\documentclass[aps,shownopacs,superscriptaddress,nofootinbib,preprint,11pt]{revtex4}
\usepackage{hyperref}
\usepackage{amsmath}
 \usepackage{multirow}
\usepackage{array}
\newcolumntype{L}[1]{>{\raggedright\let\newline\\\arraybacksslash\hspace{0pt}}m{#1}}
\newcolumntype{C}[1]{>{\centering\let\newline\\\arraybackslash\hspace{0pt}}m{#1}}
\newcolumntype{R}[1]{>{\raggedleft\let\newline\\\arraybackslash\hspace{0pt}}m{#1}}
\usepackage{float}
\usepackage{graphicx}
\usepackage{epsfig}
\usepackage{psfrag}
\usepackage{color}
\usepackage{slashed}
\usepackage{ulem}

\usepackage{amsfonts}
\usepackage{amssymb}
\usepackage{tikz}
\usepackage{tikz}
\usetikzlibrary{positioning,arrows}
\usetikzlibrary{decorations.pathmorphing}
\usetikzlibrary{decorations.markings}

\newcommand*{\be}{\begin{equation}}
\newcommand*{\ee}{\end{equation}}
\newcommand*{\bea}{\begin{eqnarray}}
\newcommand*{\eea}{\end{eqnarray}}

\newcommand{\comment}[1]{}

\newcommand{\cref}[1]{Chapter~\ref{c.#1}}

\def\beq{\begin{equation}}
\def\eeq{\end{equation}}
\def\bea{\begin{eqnarray}}
\def\eea{\end{eqnarray}}
\def\ba{\begin{array}}
\def\ea{\end{array}}
\def\bi{\begin{itemize}}
\def\ei{\end{itemize}}
\def\be{\begin{enumerate}}
\def\ee{\end{enumerate}}
\def\bc{\begin{center}}
\def\ec{\end{center}}
\def\bt{\begin{table}}
\def\et{\end{table}}
\def\btb{\begin{tabular}}
\def\etb{\end{tabular}}

\def\lsim{\raise0.3ex\hbox{$\;<$\kern-0.75em\raise-1.1ex\hbox{$\sim\;$}}}
\def\gsim{\raise0.3ex\hbox{$\;>$\kern-0.75em\raise-1.1ex\hbox{$\sim\;$}}}
	
 \preprint{
 	TIFR/TH/17-17
 }
\begin{document}

\title{Discovery prospects of a light Higgs boson at the LHC in type-I 2HDM}
\author{Disha Bhatia}
\email{disha@theory.tifr.res.in}
\address{Department of Theoretical Physics, Tata Institute of Fundamental Research, Homi Bhabha Road, Colaba, Mumbai 400 005, India}

\author{Ushoshi Maitra}
\email{ushoshi@theory.tifr.res.in}
\address{Department of Theoretical Physics, Tata Institute of Fundamental Research, Homi Bhabha Road, Colaba, Mumbai 400 005, India}

\author{Saurabh Niyogi}
\email{saurabhphys@gmail.com}
\address{Department of Physics and Astrophysics, University of Delhi, Delhi 110007, India}

\begin{abstract}
{We present a comprehensive analysis of observing a light Higgs boson in the mass range $70$ -- $110$ GeV at the 13/14 TeV LHC, in the context of the type-I two-Higgs-doublet model. 
The decay of the light Higgs to a pair of bottom quarks is dominant in most parts 
of the parameter space, except in the fermiophobic limit. Here its decay to bosons, (mainly a pair of photons), becomes important.
We perform an extensive collider analysis for the $b\bar{b}$ and $\gamma \gamma$ final states.
The light scalar is tagged in the highly boosted regimes for the $b \bar{b}$ mode to reduce the enormous QCD background.
This decay can be observed with a few thousand fb$^{-1}$ of integrated luminosity at the LHC. 
Near the fermiophobic limit, the decay of the light Higgs to a pair of photons can even be probed with a few hundred fb$^{-1}$ of integrated luminosity at the LHC.
} \end{abstract}


\maketitle

\section{Introduction}
\label{intro}
The recently discovered scalar particle at the LHC~\cite{Aad:2012tfa,Chatrchyan:2012xdj} closely resembles the Higgs boson
conjectured in the Standard Model (SM), as its measured couplings with 
the gauge bosons and fermions are in reasonable agreement 
with the SM predictions~\cite{Khachatryan:2016vau}.
However, the current measurements~\cite{Khachatryan:2016vau} still do not rule out the possibility of the observed particle belonging to an enlarged scalar sector of a beyond-the-SM scenario.  
Usually the additional scalars are considered to be heavy, 
and in some cases, they are even decoupled from the low-energy effective theory. 
However, there may exist scenarios where some of the new physics particles are lighter than the observed Higgs. We explore this possibility in the context of the two-Higgs-doublet model (2HDM) at the 13/14 TeV LHC.

The 2HDM is one of the simplest extensions of the SM with an additional scalar doublet charged under $SU(2)_L$. 
The generic structure of 2HDM induces large flavor-changing neutral 
currents (FCNCs) at the tree level and consequently faces severe constraints from the experimental data.
These FCNCs can be suppressed by imposing a discrete $Z_2$ symmetry. This classifies
2HDM into four categories: type-I, type-II, flipped and lepton specific~\cite{Branco:2011iw}. 
Any 2HDM model comprises of eight real scalar degrees of freedom. In the process of the spontaneous breaking of $SU(2)_L \times U(1)_Y$ symmetry, three out of these eight fields generate masses for $W^{\pm}$ and $Z$ bosons, leaving behind five physical scalars, namely, a light CP-even Higgs $(h)$, a heavy CP-even Higgs $(H)$, a pseudoscalar ($A$) and charged Higgs bosons ($H^{\pm}$).

The Higgs boson discovered at the LHC, being CP even~\cite{Aad:2013xqa,Chatrchyan:2013mxa}, can be identified with any one of the CP even states of the 2HDM. 
{We are interested in the scenarios where the observed Higgs corresponds to the heavier CP even scalar and $h$ is lighter than 125 GeV.}
The phenomenology of such a light Higgs has been thoroughly studied for all types of 2HDM. However, the constraints from vacuum stability, perturbativity, unitarity, electroweak precision measurements, flavor observables, and LHC Higgs searches are weakest for the type-I 2HDM~\cite{Ferreira:2012my,Celis:2013rcs,  Chang:2013ona, Dumont:2014wha, Bernon:2014nxa, Chang:2015goa,Bernon:2015wef, Cacciapaglia:2016tlr, Goncalves:2016qhh,ENBERG2017121}. We therefore focus on the type-I 2HDM for our analysis and study the discovery prospects of the light Higgs at the future runs of the LHC \footnote{The {phenomenology of such} a light CP even scalar has also been studied in the context of various supersymmetric models; see Refs.~\cite{Bhattacherjee:2015qct,Bechtle:2016kui,Cao:2013gba,Guchait:2015owa} and references therein. Also see Refs.~\cite{Gunion:2002zf,Chiang:2013ixa, Eberhardt:2013uba, Inoue:2014nva, Celis:2013ixa,Craig:2013hca, Chen:2013rba, Arhrib:2013oia,Carena:2013ooa,Bernon:2015qea} for analyses where the lighter CP even Higgs boson
was identified with the observed scalar and the remaining scalars ($H$, $H^\pm$ and $A$)
were assumed to be heavy.}. 
We choose the mass range from $70-110$ GeV to avoid decay of the observed 125 GeV Higgs to a pair of on-shell light Higgses,
i.e., $H \to hh$. 
As a result, the bounds coming from the total decay width measurement of the observed scalar~\cite{Khachatryan:2015mma}, the measurement of Higgs signal rate~\cite{Khachatryan:2016vau}, and direct decay of the observed Higgs to a pair of light Higgses, i.e., $H\to hh$~\cite{Khachatryan:2017mnf}
are irrelevant in our case
(see Refs.~\cite{Bernon:2014nxa, Aggleton:2016tdd} for an analysis with additional scalars lighter than $m_h/2$).

To study the discovery prospects of the light CP even scalar, a suitable choice 
of production and decay channels is essential. In our scenario, the light Higgs decays dominantly to
$b\bar{b}$, except in the fermiophobic limit. Here its decay to bosons (mainly photons) becomes important. We therefore examine the light Higgs decays in the $b\bar{b}$ and $\gamma \gamma$ final states at the LHC.
Note that the search for such low mass scalars decaying to diphotons has already been performed at LHC Run-1~\cite{CMS-PAS-HIG-14-037, PhysRevLett.113.171801}. 
For the diphoton channel, we consider the production of the scalar through gluon fusion and in association with gauge bosons. The production of the light scalar in association with gauge boson/top pair is considered for the $b\bar{b}$ mode. Owing to a clean environment, the diphoton final state is one of the favorite channels to search for new resonances at the LHC. In contrast, the $b\bar{b}$ state is plagued by the huge SM multijet backgrounds. 
Therefore, we consider the light Higgs in the boosted regimes for this channel, where the jet substructure techniques enable the efficient suppression of the SM backgrounds~\cite{Butterworth:2008iy, Plehn:2011tg}.

The paper is organized in the following manner.
We begin with a brief introduction to the 2HDM in Sec.~\ref{sec:2hdm}, followed by a discussion of plausible channels which can be used to probe the light Higgs at the LHC in Sec.~\ref{sec:decay}. In Sec.~\ref{sec:param}, we briefly review various constraints on the 2HDM parameter space arising from the LEP and LHC measurements, in the context of Type-I 2HDM. A dedicated collider analysis of the light Higgs in the allowed parameter space at the LHC is performed in Sec.~\ref{LHC_future}. Finally in Sec.~\ref{sec:discuss}, we summarize our results. Further in appendix~\ref{sec:loop}, we discuss the light Higgs couplings to diphotons. In Appendix~\ref{sec:chargedHiggs}, the implications of the light charged Higgs boson on our results is analyzed, and in Appendix~\ref{sec:fatjet}, the tagging methods used to reconstruct boosted objects are discussed.
Finally, in Appendix~\ref{sec:csplot}, we tabulate the behaviour of the total cross section of the selected modes with respect to the 2HDM parameters. 

\section{2HDM: a brief review}
\label{sec:2hdm}
The $Z_2$-symmetric 2HDM Lagrangian with two $SU(2)_{L}$ Higgs doublets ($\Phi_{1}$ and $\Phi_{2}$)\footnote{Under $Z_2$ transformation, $\Phi_1 \to \Phi_1$ and $\Phi_2 \to - \Phi_2$ } can be parametrized as~\cite{Branco:2011iw}:
 \begin{equation}
  \mathcal{L}_{\text{2HDM}}=~\left(D_{\mu}\Phi_1\right)^\dagger\,D^{\mu} \Phi_1 + \left(D_{\mu} \Phi_2\right)^\dagger\, D^{\mu} \Phi_2  + \mathcal{L}_{\text{Yuk}} (\Phi_{1}, \Phi_{2}) - V(\Phi_{1}, \Phi_{2})\;,
\label{eqn:Lag}
 \end{equation}
where $\mathcal{L}_{\text{Yuk}}$ represents the Yukawa interactions and $V(\Phi_{1}, \Phi_{2})$
is the scalar potential given as
\begin{eqnarray}
V(\Phi_{1}, \Phi_{2})  &=&  m_{11}^2\,\Phi_1^\dagger \Phi_1 +\frac{\lambda_1}{2}\,(\Phi_{1}^\dagger \Phi_{1})^2
                             + m_{22}^2\,\Phi_{2}^\dagger \Phi_{2} 
                             + \frac{\lambda_2}{2} \,(\Phi_{2}^\dagger \Phi_{2})^2 - \left[ m^2_{12} \Phi_{1}^{\dagger}\Phi_{2} + h.c \right] \nonumber \\
                        && + \lambda_3\,\Phi_{1}^\dagger \Phi_{1}\,\Phi_{2}^\dagger \Phi_{2}  
                           + \lambda_4 \,\Phi_{1}^\dagger \Phi_{2} \,\Phi_{2}^\dagger \Phi_{1}
                         - \left[ \frac{1}{2} \lambda_5 \left(\Phi_{1}^\dagger \Phi_{2}\right)^2 + H.c. \right]\;.
\label{eq:pot}
\end{eqnarray}
Here $m_{12}$ is the soft $Z_2$ symmetry-breaking parameter. Note that in our analysis, we have assumed $V(\Phi_{1},\Phi_{2})$ to be invariant under CP (i.e., charge and parity transformations) and consequently the parameters of the scalar potential are real. 
The spontaneous breaking of $SU(2)_L \times U(1)_Y$ symmetry results in five physical scalar fields $h$, $H$, $A$ and $H^\pm$ (with masses $m_h$, $m_H$, $m_A$ and $m_{H^{\pm}}$, respectively) and three Goldstone bosons $G$ and $G^\pm$, which appear as the longitudinal modes for $Z$ and $W^\pm$ bosons. The mass spectra of the particles are obtained by minimizing the scalar potential $V(\Phi_1,\Phi_2)$ in Eq.~(\ref{eq:pot}).

The doublets in terms of the physical fields and the Goldstone bosons {can be expressed as:}   
\begin{eqnarray}
\Phi_1 &=& \begin{pmatrix}
               G^+\cos\beta + H^+ \sin\beta \\
              \frac{1}{\sqrt{2}} \left[h \sin\alpha - H \cos\alpha 
               +i \left( G \cos\beta + A \sin\beta  \right)  + v_1 \right]
         \end{pmatrix} \;,
         \nonumber \\
\Phi_2 &=& \begin{pmatrix}
               G^+\sin\beta - H^+ \cos\beta \\
              \frac{1}{\sqrt{2}} \left[-h \cos\alpha - H \sin\alpha 
               +i \left( G \sin\beta - A \cos\beta  \right)  + v_2 \right]
         \end{pmatrix} \;,
         \label{eqn:doublets}
\end{eqnarray}
where $\alpha$ and $\beta$ are the rotation angles which diagonalize the mass matrices for the 
neutral CP even Higgs and the charged Higgs/CP odd Higgs {respectively.} 
The parameters of the scalar potential ($m_{11} , m_{22} , \lambda_i$ )  can be expressed
in terms of the rotation angles ($\alpha, \beta$), the $Z_2$ symmetry breaking
parameter ($m_{12}$), and the masses of the scalars ($m_h$, $m_H$, $m_A$, $m_H^\pm$) as~\cite{Branco:2011iw}:

\begin{eqnarray}
m_{11}^2 &=& \frac{1}{4}\bigg( m_h^2 + m_H^2 - 4 m_{12}^2 \tan\beta + \left( m_H^2 - m_h^2\right) \sec\beta \cos(2\alpha-\beta) \bigg)\;,\\
m_{22}^2 &=& \frac{1}{4}\bigg( m_h^2 + m_H^2 - 4 m_{12}^2 \cot\beta + \left( m_H^2 - m_h^2\right) \csc\beta \sin(2\alpha-\beta) \bigg)\;,\\
\lambda_1 &=& \frac{1}{2v^2}\sec^2\beta \left( m_h^2 + m_H^2 + (m_H^2 - m_h^2) 
\cos 2\alpha 
              -2 m_{12}^2 \tan\beta \right)\;,\\
\lambda_2 &=& \frac{1}{2v^2} \csc^2\beta 
\left( m_h^2 + m_H^2 - (m_H^2 - m_h^2) 
\cos 2\alpha -2 m_{12}^2 \cot\beta \right)  \;,\\
\lambda_3 &=& \frac{1}{v^2} \csc 2\beta 
\left( -2 m_{12}^2 + (m_H^2 - m_h^2) \sin 2\alpha + 2 m^2_{H^{\pm}} \sin 2 \beta   \right) \;, \\
\lambda_4 &=& \frac{1}{v^2}\left( m_A^2 -2 m^2_{H^{\pm}} + m_{12}^2 \csc \beta \sec \beta \right)   \;,   \\
\lambda_5 &=& \frac{1}{v^2} \left( m^2_{12} \csc \beta \sec \beta - m_A^2  \right) \;, \\
v_1 &=& v \cos \beta \quad \text{and} \quad v_2 \,\,= \,\, v \sin \beta \;.
\end{eqnarray} 

%
 The couplings $\lambda_{i}$ ($i=1,5$) are constrained by the perturbativity, vacuum stability ~\cite{Nie:1998yn}, and unitarity~\cite{Akeroyd:2000wc} bounds,
which in turn, restrict the allowed values of the scalar masses for a given value of $\alpha$ and $\beta$ \cite{Branco:2011iw,Bhattacharyya:2015nca}.  
The masses of the additional scalars also get constrained from the well measured flavour and electroweak observables~\cite{{ALEPH:2005ab,Mahmoudi:2009zx,Aoki:2009ha}}. The combined effect of these constraints on the 2HDM parameter space is discussed in Appendix.~\ref{sec:chargedHiggs}, in the context of type-I 2HDM. Note that the free parameter $\alpha$ remains unaffected 
after imposition of above constraints (see Fig. \ref{fig:chargedHiggs} in appendix.~\ref{sec:chargedHiggs}).

In our analysis, we identify the heavier CP even Higgs with the discovered scalar by fixing $m_H=125$ GeV and study the phenomenology of the light CP even scalar $h$. At this stage we have following free parameters --- $\alpha$, $\beta$, $m_{H^\pm}$, $m_A$, $m_{12}$ and $m_h$. However, we confine ourselves to that part of the allowed parameter space, where the masses of the charged and pseudoscalar Higgs bosons are heavy i.e. $\mathcal{O}(500)$ GeV and do not affect our analysis. The $Z_2$ breaking parameter in this case becomes irrelevant for the light Higgs phenomenology and can be suitably chosen
to have any value less than $100$ GeV (see appendix.~\ref{sec:chargedHiggs}).  
Note that although we have chosen the charged Higgs to be heavy for most of our analysis, we do analyze the implications of having a low-mass charged scalar in Appendix.~\ref{sec:chargedHiggs}. 

We now discuss the couplings of the scalar particles with fermions and gauge bosons. 
In the type-I 2HDM, fermions couple only to one of the doublets i.e. $\Phi_2$
and $\mathcal{L}_{\rm Yuk}$ is given as
\begin{eqnarray}
\mathcal{L}^{\rm Type-I}_{\rm Yuk}  
                         &= &\overline{{Q_{L}}}\, \mathcal{Y}^d \Phi_2 d_R 
                          +  \overline{Q_L} \mathcal{Y}^u \Phi^c_2 u_R 
                          +  \overline{Q_L} \mathcal{Y}^e \Phi^c_2 e_R + h.c. \;,\nonumber \\
                          && \nonumber \\
&= & -\sum_{f=u,d,\ell}\frac {m_f}{v} \left( \xi_h^f \overline{f}f\,h + \xi_H^f  \overline{f}f\,H 
      -i\xi_A^f \overline{f}\gamma_5f\,A \right) \nonumber \\
  &-&   \frac{\sqrt{2}V_{ud}}{v}  \overline{u} \left( {m_u} \xi_A^u P_L + {m_d} \xi_A^d P_R  \right) d \, H^+
         - \frac{\sqrt{2} {m_{\ell}}}{v} \xi_A^\ell \overline{\nu} P_R \ell \, H^+ + h.c. \;,                             
\label{eq:type1}
\end{eqnarray}
where $\mathcal{Y}^{u,d,e}$ are $3\times 3$ Yukawa matrices, $V_{ud}$ is the Cabibbo-Kobayashi-Maskawa matrix element, $m_f$ is the mass of a fermion ($f$) and
\begin{eqnarray}
\xi_h^{u,d,\ell} = \cos\alpha/\sin\beta\;,\; \xi_H^{u,d,\ell} = \sin\alpha/\sin\beta \;,\;  
\xi_A^u = \cot{\beta}\;,\; \xi_A^{d,\ell} = -\cot{\beta}\;.
\label{eqn:ffH}
\end{eqnarray}
We list some of the couplings of gauge bosons with scalars that are relevant for our analysis ({see Refs.~\cite{Gunion:1989we},~\cite{Barger:1987nn} for a complete list}):
\begin{eqnarray}
\mathcal{L}_{\rm Gauge-int} &=& 
\frac{m_Z^2}{v} \xi_h^V  Z_\mu Z^\mu h + \frac{m_Z^2}{v} \xi_H^V  Z_\mu Z^\mu H  
+ 2\frac{m_W^2}{v}\xi_h^V  W_\mu W^{\mu} h
+ 2 \frac{m_W^2}{v}\xi_H^V  W_\mu W^{\mu} H   \nonumber \\
 &+&  \frac{\alpha_{em}}{8 \pi v}\; \xi^{\gamma}_{h} h\; F_{\mu \nu} F^{  \mu \nu}  
 \;\;+\frac{\alpha_{em}}{8 \pi v}\; \xi^{\gamma}_{H} \;H F_{\mu \nu} F^{ \mu \nu} \;,
\label{eq:VVH}
 \end{eqnarray}  
where 
\begin{equation}
\xi_h^{V} = \sin({\beta-\alpha}) \quad \;, \quad \xi_H^{V} = \cos({\beta-\alpha})\,,
\label{eqn:Vh}
\end{equation}
and the expressions for $\xi_h^{\gamma}$ and $\xi_H^{\gamma}$ are listed in Appendix~\ref{sec:loop}.
Based on the above couplings of the scalars with fermions and gauge bosons, two 
interesting limits arise:
\begin{enumerate}
\item Alignment limit ($\alpha \to \beta$): Here
the couplings of the heavier CP even Higgs exactly match those of the SM Higgs.  

\item Fermiophobic limit ($\alpha \to \pi/2$) : In this limit, the tree-level couplings of the light Higgs with fermions ($\xi^{f}_{h}$) vanish [see Eq.~(\ref{eqn:ffH})] and its loop-induced couplings with fermions are also
negligible. The light Higgs in this case behaves as a fermiophobic scalar. 

\end{enumerate}

We shall see later that 
these limits have interesting implications in our analysis. 
As an aside, note that the condition $\alpha \equiv \beta \to \pi/2$ corresponds to the case where 
the alignment and fermiophobic limits occurs simultaneously. In this case, the couplings of the light Higgs with both fermions and gauge bosons vanish [see Eqns.~\ref{eqn:ffH}, \ref{eqn:Vh}] and the type-I 2HDM maps to the inert 2HDM model.

After discussing the couplings of the light Higgs boson, we are now in a position to predict its phenomenological consequences.  In the next section, we identify the promising channels which could be useful in probing the 
light Higgs at the LHC. Keeping a large QCD background in mind, a suitable choice of the production channel and decay mode would be essential for the discovery of a light scalar like the SM Higgs.

\section{Promising channels to explore at the LHC}
\label{sec:decay}

\begin{figure} [h]
 \begin{center}
\includegraphics[width=0.45\textwidth,height=6.0cm]{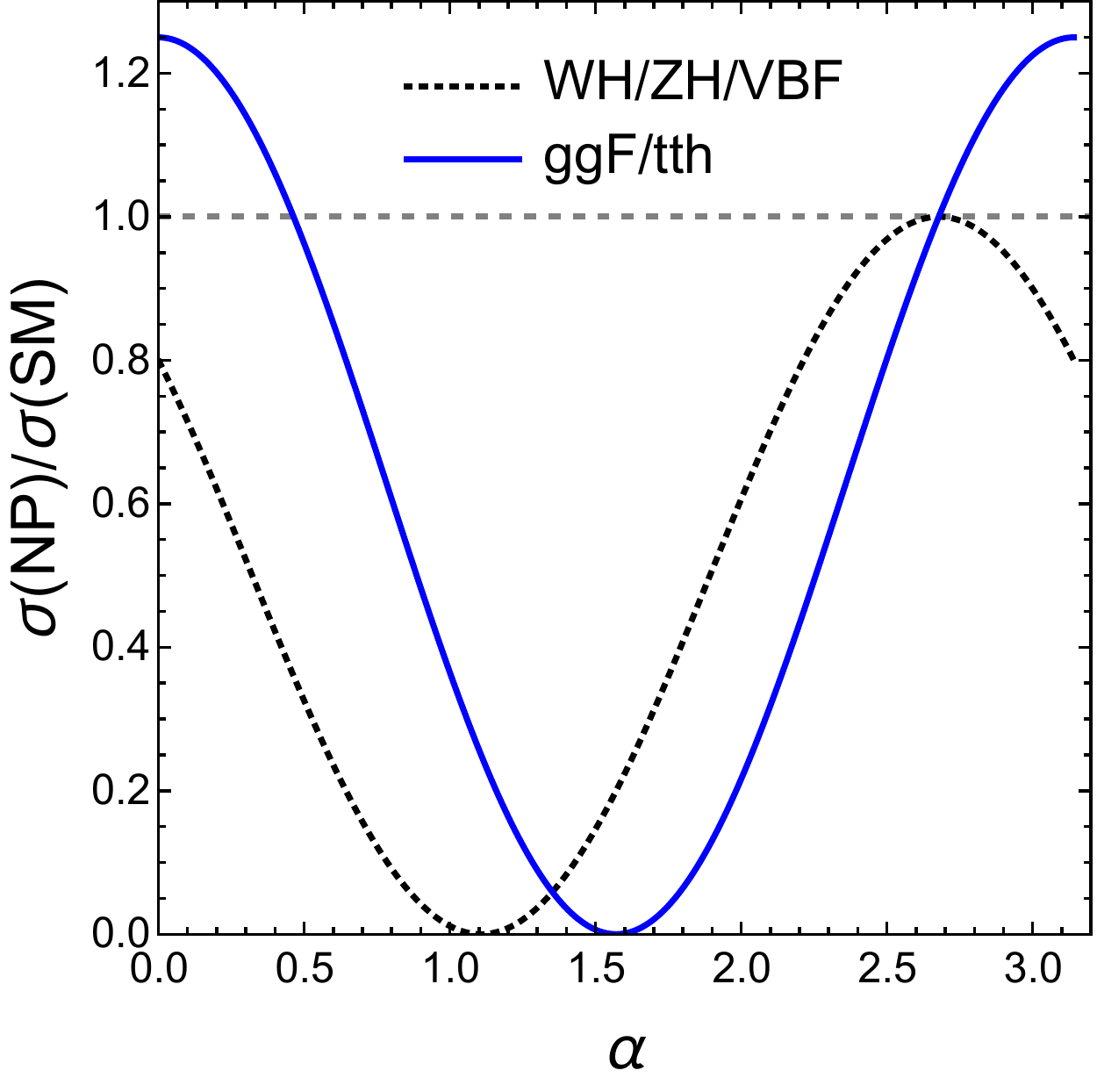}\quad
\includegraphics[width=0.45\textwidth,height=6.0cm]{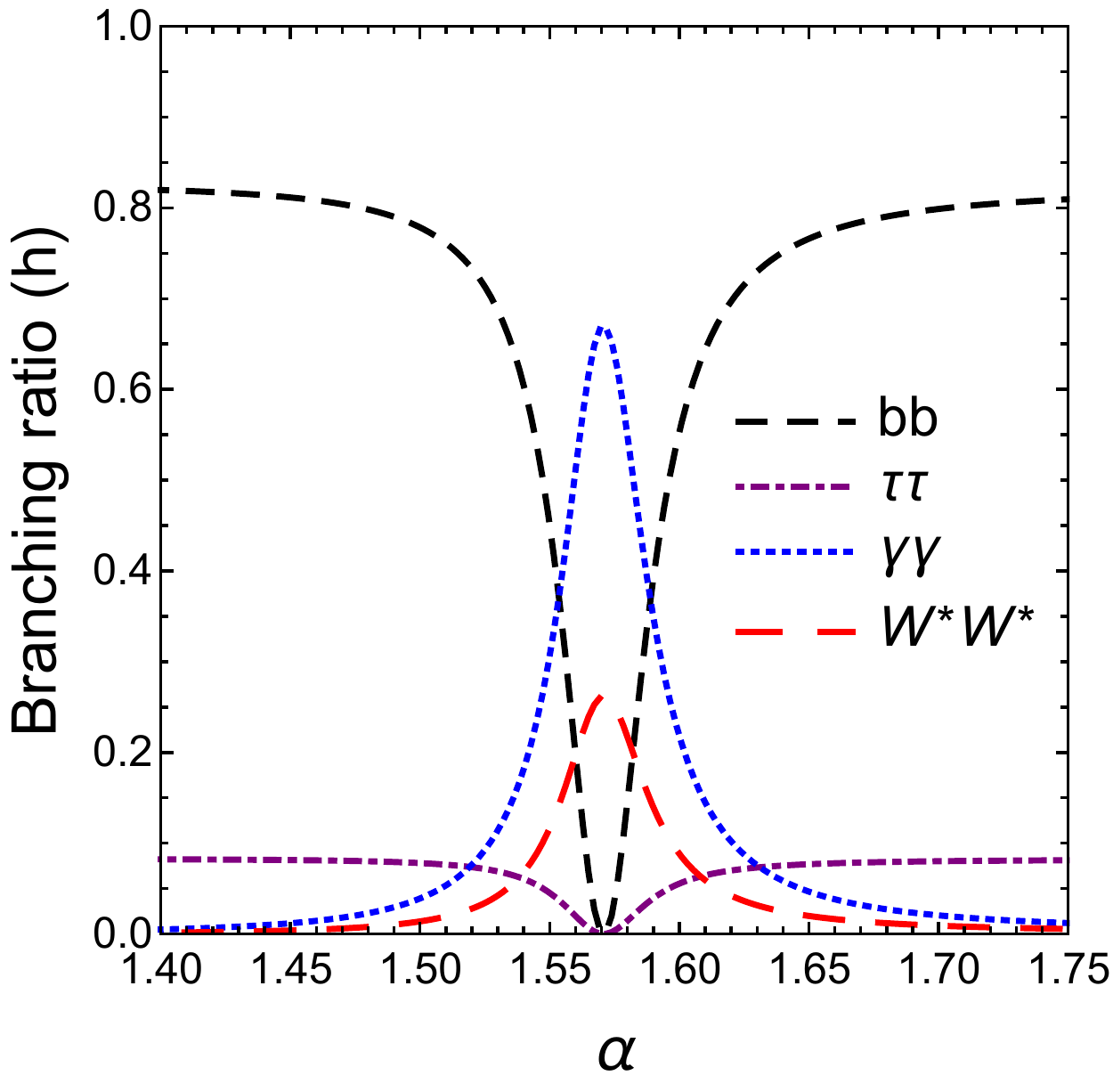}
\caption{\label{fig:1} A representative plot for $m_h = 80$ GeV and $\tan \beta =2$. The left panel illustrates the variation of the ratio of the cross sections of the light Higgs ($h$) and the SM-like Higgs ($h_{\text{SM}}$) with $\alpha$. The right panel shows the branching ratios of the light Higgs as a function of $\alpha$ (the range of $\alpha$ is restricted near $\pi/2$ to signify the behavior around the fermiophobic limit). Note that the $Vh$/VBF
 and ggF/$t\bar{t}h$ production modes are able to probe a similar parameter space. In a similar fashion, the variations of the $b\bar{b}$ and $\tau \bar{\tau}$ decay modes are identical with $\alpha$.}
 \end{center}
 \end{figure}

The light Higgs boson, just like the SM Higgs, can be produced at the LHC via gluon fusion (ggF), vector-boson fusion (VBF), and in association with SM gauge bosons (V$h$), as well as with a top pair ($t\bar{t}h$). The ratio of the production cross section of the light Higgs and that of the SM-like Higgs ($h_{SM}$) as a function of $\alpha$ is plotted in the left panel of Fig.~\ref{fig:1}. Here, by ``SM-like Higgs" we mean a hypothetical scalar whose couplings are exactly same as those of the SM Higgs but whose mass is equal to that of the light Higgs i.e. $m_{h_{SM}} = m_h$. Note that $m_h$ is chosen to be 80 GeV in Fig.~\ref{fig:1} for illustrative purposes. The gluon fusion as well as the $t\bar{t}h$ production cross section of the light Higgs in the type-I model scale as $(\xi^{f}_{h})^2$ with respect to the SM-like Higgs.
Similarly, the cross sections for the light Higgs produced in association with gauge bosons or through vector-boson fusion scale as $(\xi^{V}_{h})^2$. Therefore, the ordinate in Fig.~\ref{fig:1} essentially shows the variation of $(\xi^{i}_{h})^2$ against $\alpha$. 
The scaling has been illustrated for $\tan \beta = 2$ in Fig.~\ref{fig:1}. It can be seen from
Eqn.~(\ref{eqn:ffH}) and (\ref{eqn:Vh}) that for large values of $\tan \beta$, all production channels scale identically
as $\cos^2 \alpha$.


The right panel of Fig.~\ref{fig:1} represents various branching fractions of the light Higgs again as a function of $\alpha$. In most of the parameter space, the light Higgs decays dominantly to a pair of bottom quarks. However, near $\alpha \to \pi/2$ (the fermiophobic limit), it decays maximally to a pair of gauge bosons. 
We therefore choose $b\bar{b}$ and $\gamma\gamma$ as the light Higgs decay modes for our analysis. 
We must stress that the branching ratio of $h$ to a $\tau$ pair is also significant ($\sim 10\%$). Since the parameter space probed
by it is similar to that of $b\bar{b}$,
we restrict ourselves to the analysis of $b\bar{b}$ in this manuscript.

Now our task is to determine the suitable production mode for a light Higgs decaying to a pair of bottom quarks and photons. 
Note that analyzing $b\overline{b}$ channel in the ggF or VBF mode is challenging due to the presence of large QCD background. However, the presence of a lepton(s) in addition to the $b\bar{b}$ in $Vh$ or $t\bar{t}h$ production modes could help to suppress these backgrounds. Hence, we choose light Higgs production in association with a $W$ boson and top pair for $b\bar{b}$ analysis.\footnote{$Zh$ production mode is neglected as leptonic branching ratio in case of $Z$ is smaller than $W$ and the parameter space probed by $Wh$ and $Zh$ are exactly the same.} On the other hand, the diphoton channel is one of the cleanest probes for discovering new resonances at the LHC. This channel is also comes with the additional advantage of enhanced sensitivity near the fermiophobic limit in the type-I 2HDM.     
In this limit, i.e., $\alpha \to \pi/2$, the decay of $h$ to $\gamma \gamma$ becomes prominent and can only be probed through the $Vh$/VBF production mode as shown in Fig.~\ref{fig:1}. 
We have considered only the $Wh$ process in our analysis as the parameter space probed by VBF and $Zh$ are exactly the same as that of $Wh$. 
The diphoton channel can also be used to probe regions away from the fermiophobic limit 
through ggF/$t\bar{t}h$ production mode. 
Since the production cross section of $t\bar{t}h$ is roughly 100 times smaller than that of ggF, we have not considered this for the diphoton analysis. 
To summarize, we have chosen the following channels\footnote{The behavior of the total cross section with respect to $\alpha$ and $\beta$
corresponding to the four selected channels is discussed in Appendix~\ref{sec:csplot}.} for probing the light Higgs at the LHC:  \\
{\rm Channel 1}: $p p \to h \to \gamma\gamma$. \\
 {\rm Channel 2}: $p p \to W h \to W \gamma\gamma$. \\
 {\rm Channel 3}: $p p \to W h \to b\bar{b}$. \\
 {\rm Channel 4}: $p p \to t\bar{t} h \to t\bar{t} b\bar{b}$.\\
The phenomenological consequences of these channels will be examined in Sec.~\ref{LHC_future}. 
\section{Experimental constraints}
\label{sec:param}

In this section, we discuss the experimental constraints on the 2HDM parameter space i.e. $\alpha$, $\tan\beta$ and $m_h$ from the observed Higgs signal strength measurements and 
the direct searches of light scalars at LEP and LHC. 
In our analysis, we have varied $\alpha$ in its full range i.e., $[0: \pi]$ and 
$\tan\beta$ in the restricted range of $[1:10]$. 
While the lower value of $\tan\beta$ is chosen to account for 
the constraints from the flavor observables (as discussed in Appendix.~\ref{sec:chargedHiggs}, the higher value is restricted to 10 for interesting phenomenology.\footnote{With an increase in $\tan\beta$, the couplings of the light Higgs in the type-I 2HDM with fermions decrease, and 
for gauge bosons, they become independent of $\beta$.}
The organization of this section is as follows. In Secs.~\ref{sec:LHC_bounds} and \ref{sec:LEP} we discuss the individual constraints from the signal strength measurements and 
the direct searches for light scalars, respectively. Towards end of Sec.~\ref{sec:LEP}, the combined effect of the above constraints on the parameter space is presented.

\subsection{LHC constraints: Signal strength measurements of the 125 GeV Higgs}
\label{sec:LHC_bounds}

Since we have identified the heavier CP even Higgs with the observed Higgs boson, its couplings with fermions and 
gauge bosons -- which are different from that of the SM by the factors $\xi_H^f$, $\xi_H^V$, $\xi^{\gamma}_{H}$, get constrained by the signal strength measurements~\cite{Khachatryan:2016vau}. 
If the observed Higgs is produced through channel $i$ and decays 
to $j$, then the signal strength ($\mu_j^i$) (assuming narrow-width approximation) is
defined as \cite{Dawson:2013bba,Khachatryan:2016vau}
\begin{eqnarray}
\mu^i_j = \frac{\sigma(i \to H)}{\sigma(i \to H_{\rm SM})} \times \frac{{\rm BR}( H \to j)}{{\rm BR}( H_{\rm SM} \to j)} \;= \;
\xi_H^{\text{prod},i} \times \xi_H^{\text{decay},j} \frac{\Gamma_{H_{\text{SM}}}^{\text{tot}}}{\Gamma_{H}^{\text{tot}}}\;,
\end{eqnarray}  
where
\begin{eqnarray}
\xi_H^{\text{prod},i}   = \frac{\sigma(i \to H)}{\sigma(i \to H_{\rm SM})}\;,
\quad \xi^{\text{decay},j}_H   = \frac{{\Gamma}( H \to j)}{{\Gamma}( H_{\rm SM} \to j)} \;,
\quad  \Gamma_H^{\text{tot}} =   \sum_k \xi^{\text{decay},k}_H \Gamma_H^{k,\text{SM}}  \; \nonumber.
\end{eqnarray} 
In Table~\ref{tab:tab1} we list the production and decay scaling factors for the observed Higgs. Note that these factors are exact only at the leading order. However, the deviations after including the higher-order corrections are small~\cite{Cacciapaglia:2016tlr} and hence are neglected in the analyses.

\begin{table}[h]
\begin{center}
\def\arraystretch{1.5}
\begin{tabular}{|c|cc||c|ccc |} \hline
 Production       & $ggF/t\bar{t}H$  & VBF$/VH$ & Decay &$f\bar{f}$  & $VV^*$ & $\gamma \gamma$ \\ \hline \hline
$\xi_H^\text{prod}$ &   $(\xi^{f}_{H})^{2}$ & $(\xi^{V}_{H})^{2}$ &
 $\xi_H^{\text{decay}}$ & 
$(\xi^{f}_{H})^{2}$ &    
$(\xi^{V}_{H})^{2}$  & $(\xi^{\gamma}_{H})^{2}$     \\ 
\hline 
\end{tabular}
\end{center}
 \caption{\label{tab:tab1} Scaling factors for the production and decay processes. See Eqs.~(\ref{eqn:ffH}),~(\ref{eq:VVH}) and~(\ref{eqn:gammaH}) for the definitions of $\xi^{f}_{H}$, $\xi^{V}_{H}$ and $\xi^{\gamma}_{H}$ respectively.}
\end{table}
\begin{table}[h]
\begin{center}
\begin{tabular}{|c|c||c|c|} \hline
 Signal Strength        & ATLAS-CMS ($7-8$ TeV)  &  Signal Strength        & ATLAS-CMS ($7-8$ TeV)\\
 ($\mu^\text{ggF}_j$)$^{\rm exp}$            &   (combined) & ($\mu^\text{VBF}_j$)$^{\rm exp}$            &   (combined)\\ \hline \hline
 $\mu^{\text{ggF}}_{\gamma\gamma}$ & $1.10^{+0.23}_{-0.22}$ & $\mu^{\text{VBF}}_{\gamma\gamma}$ & $1.3^{+0.5}_{-0.5}$\\ \hline
 $\mu^{\text{ggF}}_{ZZ}$           & $1.13^{+0.34}_{-0.31}$ & 
 $\mu^{\text{VBF}}_{ZZ}$           & $0.1^{+1.1}_{-0.6}$  \\ \hline 
 $\mu^{\text{ggF}}_{WW}$           & $0.84^{+0.17}_{-0.17}$ 
 &  $\mu^{\text{VBF}}_{WW}$        & $1.2^{+0.4}_{-0.4}$\\ \hline 
  $\mu^{\text{ggF}}_{\tau \bar{\tau}}$    & $1.0^{+0.6}_{-0.6}$ 
 &  $\mu^{\text{VBF}}_{\tau \bar{\tau}}$  & $1.3^{+0.4}_{-0.4}$\\ \hline 
\end{tabular}
\end{center}
 \caption{\label{tab:bounds_sig} The combined measured values of $(\mu^i_j)^{\rm exp}$ from ATLAS and CMS using 7 and 8 TeV data~\cite{Khachatryan:2016vau}, used in our analysis. The allowed regions in the parameter space are determined by allowing individual $\mu^i_j$ predicted in the type-I 2HDM to lie within $\pm 2 \sigma$ from the central values of the measured signal strengths i.e. $(\mu^i_j)^{\rm exp}$. }
\end{table}

\begin{figure} [h]
 \begin{center}
 \includegraphics[totalheight=7.0cm]{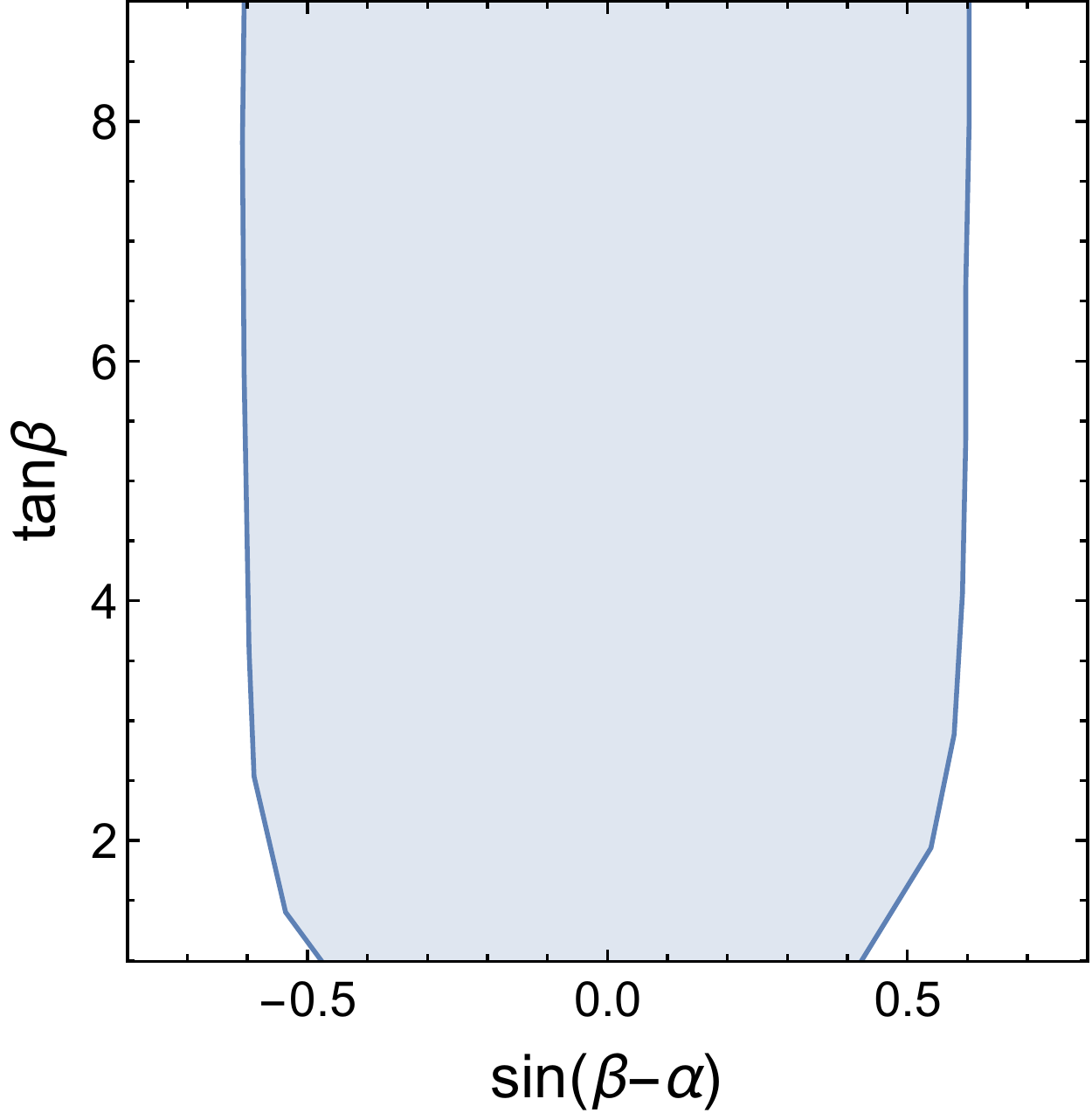}
 \caption{\label{param_lhcsignal} The allowed region (shaded in blue) is determined by allowing individual $\mu_i^j$ predicted
in the type-I 2HDM to lie within $\pm 2\sigma$ of the central values of $(\mu^i_j)^{\rm exp}$ obtained from the combined
ATLAS and CMS 7 and 8 TeV data~\cite{Khachatryan:2016vau}. The signal strengths considered for the analysis are listed in Table~\ref{tab:bounds_sig}.
Note that we have not employed the $\chi^2$ minimization technique in our analysis for determining the allowed regions of the parameter space.}
 
 \end{center}
 \end{figure}

The measured signal strengths i.e. $(\mu^i_j)^{\rm exp}$ used in our analysis are listed in Table~\ref{tab:bounds_sig}.
Note that these measurements do not constrain the mass of the light Higgs due to the absence of $H\to hh$ the decay mode as we have considered $m_h > m_H/2$ in our analysis. Hence the parameters that are constrained by Higgs signal strength measurements are $\alpha$ and $\beta$. 
In Fig.~\ref{param_lhcsignal}, we present the allowed regions in the $\left(\sin(\beta - \alpha) \rm{,} \tan \beta\right)$ plane after incorporating constraints from the Higgs signal strength measurements.
These regions are determined by allowing individual $\mu^i_j$ predicted in the type-I 2HDM to lie within $\pm 2 \sigma$ of the central values of $(\mu^i_j)^{\rm exp}$ obtained from the combined ATLAS and CMS 7 and 8 TeV data~\cite{Khachatryan:2016vau}. However, we must mention that we have not employed the $\chi^2$ minimization technique while deriving such allowed regions using data.

We now discuss the qualitative features of Fig.~\ref{param_lhcsignal}. The constraints from the Higgs signal strength measurements force us to remain close to the alignment limit
as the heavier CP even Higgs here behaves exactly like the SM Higgs.
In Fig.~\ref{param_lhcsignal}, one could notice that 
for $\tan \beta \approx 1 $, negative values of $\sin(\beta - \alpha)$ are slightly less constrained than positive ones. In this region, $\sin(\beta - \alpha) > 0$ implies $\alpha < \pi/4$ and $\sin(\beta - \alpha) < 0$ implies $\alpha > \pi/4$. 
As a result, the Yukawa couplings of the SM-like Higgs which scale as ($\xi^{f}_{H}$) decrease for increasing positive values of $\sin(\beta - \alpha)$. Therefore, the signal strength $\mu^{ggF}_{j}$ (which depends on $\xi_H^f$) drops quickly below the allowed range for positive values of $\sin(\beta - \alpha)$, making this region relatively more constrained. 
For larger values of $\tan\beta$, {$\sin(\beta-\alpha)$ is approximately equal to $\cos\alpha$. Hence,
the allowed region in Fig.~\ref{param_lhcsignal} becomes symmetric in $\sin(\beta-\alpha)$ as well as independent of $\tan\beta$. Although, we have plotted effect of signal strength constraints in the ($\sin(\beta-\alpha)$, $\tan\beta$) plane, this can be easily translated to the ($\alpha$, $\tan\beta$) plane. The net effect is only to restrict the allowed range of $\alpha$ to be less than $\pi$ for a given value of $\tan \beta$.


\subsection{Light Higgs direct search bounds} 
\label{sec:LEP}
Note that as the center-of-mass energy at LEP was limited to 209 GeV, the production cross section of the light Higgs
for the heavier masses faced severe phase-space suppression. As a result, these masses are less constrained by the LEP data. 
In the right panel of Fig.~\ref{fig:sel}, we project the LEP bounds listed in the left panel onto the allowed regions at 95$\%$ C.L. in the $(\alpha,\tan\beta)$ plane
for $m_h = 90$ GeV (pink) and $m_h=100$ GeV (blue) for illustrative purposes. 
Note that the LEP constraint -- just like the Higgs signal strength -- restricts the allowed range of $\alpha$ to be less than $\pi$, for a given $\tan\beta$ and $m_h$.
{We must mention that the Tevatron also searched for such a light Higgs in the $Vh$ production mode~\cite{Aaltonen:2013js}. However, the Tevatron bounds are much less stringent than LEP and hence are not considered in the analysis. }

LEP has also searched for a CP odd scalar in the process $e^{+}e^{-} \to h A$~\cite{LEPHiggsWorking:2001ab,ref1}. This search is complimentary to  $e^{+}e^{-} \to h Z$ 
as the former depends on $\cos^2(\beta-\alpha)$
and the latter on $\sin^2(\beta-\alpha)$. The null results in both production modes significantly constrain both $\sin(\beta-\alpha)$ and $\cos(\beta-\alpha)$
and require them to be much less than unity.
If both $h$ and $A$ are light at the same time such that $m_A+m_h < 209$ GeV, 
then the combined direct search constraints of $h$ and $A$ rule out a significant part of the parameter space including the regions which satisfy the alignment limit.
Therefore, our choice of demanding a heavy pseudoscalar is in sync with the requirement of a light Higgs. 

Both the ATLAS and CMS collaborations have searched for additional light scalars in the diphoton final state~\cite{CMS-PAS-HIG-14-037, PhysRevLett.113.171801, CMS-PAS-HIG-17-013}. While CMS has placed $95\%$ C.L. upper bounds on the total cross section for a light scalar decaying to $\gamma \gamma$ for the production modes ggF$+t\bar{t}h$ and VBF$+Vh$. On the other hand, ATLAS instead provides an inclusive bound for the combination of all of the production modes.
To understand the effect of these measurements on the parameter space of the 2HDM, let us note the behavior of the
total cross section of the light Higgs decaying to a pair of photons.
\begin{figure} [t]
\begin{center}
 \includegraphics[height=6.0cm]{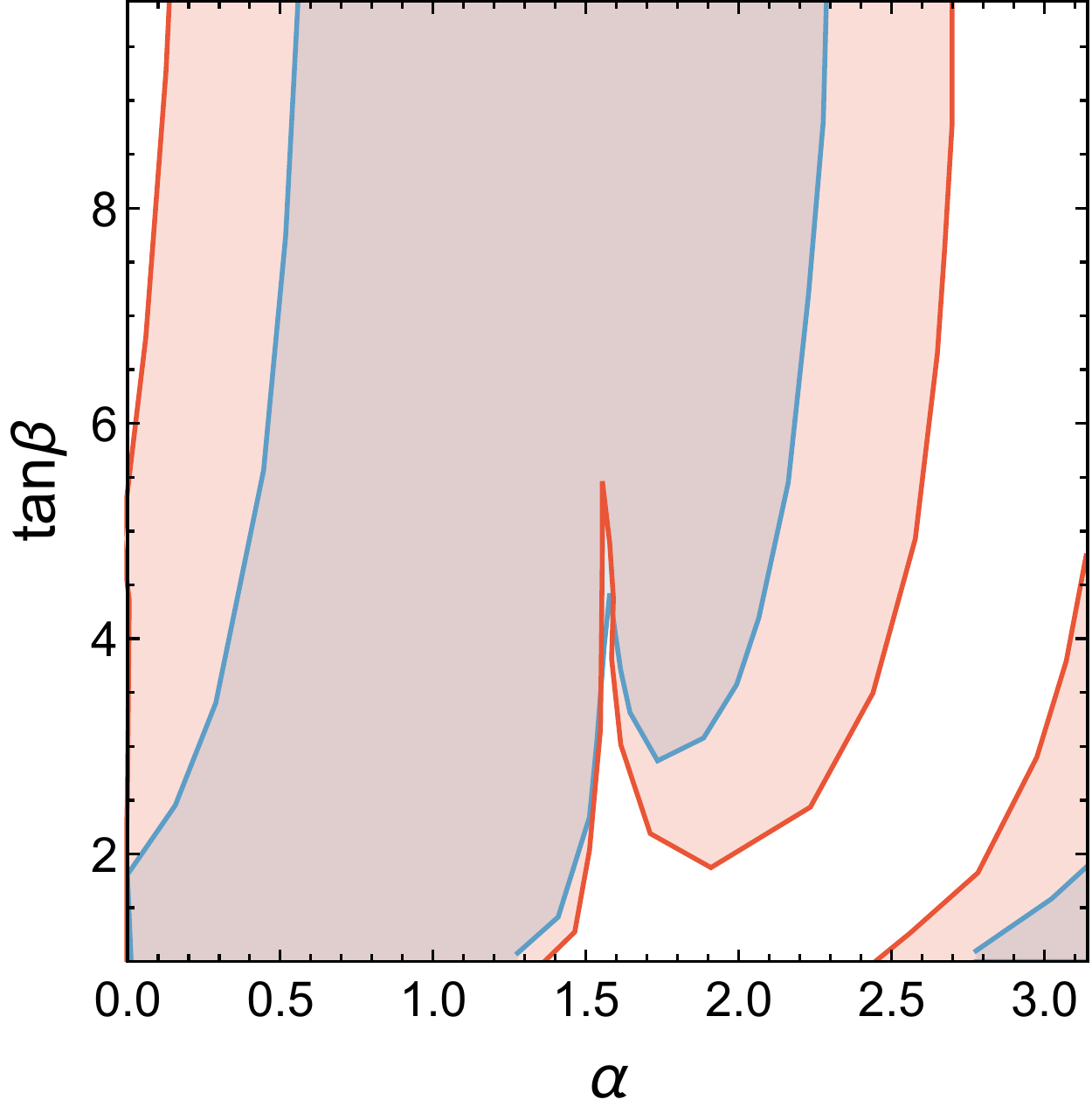}
\caption{ We demonstrate the allowed regions by incorporating constraints only from the light scalar searches at the LHC~\cite{CMS-PAS-HIG-14-037, PhysRevLett.113.171801,CMS-PAS-HIG-17-013} in the ($\alpha$, $\tan\beta$) plane for $m_h=90$ GeV (pink band) and 100 GeV (blue band). The masses have been chosen for illustrative purposes as before.}
\label{fig:LHC-8}
 \end{center}
 \end{figure} 
We now know that the light Higgs branching ratio to a pair of photons is large near the fermiophobic limit and could be probed in the VBF$+Vh$ production mode. 
However, in this case, the total cross section, i.e., $\sigma \times {\rm BR}$ is large only for smaller values of $\tan\beta$ and tends to zero for larger values of $\tan\beta$ (see Table~\ref{tab:crosssection-behaviour} in Appendix~\ref{sec:csplot}). 
For regions away from the fermiophobic limit, although the branching ratio of the light Higgs to a pair of photons is not large, 
this decay could still be probed in the ggF mode owing to its large production cross section. 

The effect of the LHC direct detection constraints~\cite{CMS-PAS-HIG-14-037, PhysRevLett.113.171801,CMS-PAS-HIG-17-013} are displayed in Fig.~\ref{fig:LHC-8}, where we plot the allowed parameter space in the ($\alpha,\tan\beta$) plane for $m_h=90$ GeV (pink band) and 100 GeV (blue band). The masses have been chosen for illustrative purposes as before. 
The combined bounds from ATLAS and CMS near the fermiophobic limit are sensitive only to the VBF$+Vh$ production mode, where the total cross section is large for smaller $\tan\beta$ values. Consequently, this region gets severely constrained and results in a wedge-like exclusion around $\alpha \approx \pi/2$ as can be seen in Fig.~\ref{fig:LHC-8}.
For regions away from the fermiophobic limit, the combined constraints from ATLAS
and CMS~\cite{CMS-PAS-HIG-14-037, PhysRevLett.113.171801, CMS-PAS-HIG-17-013} are far more stringent for $m_h=100$ GeV than 90 GeV, hence rule out a significant part of the parameter space for the same.

\begin{figure} [t]
 \begin{center}
 \includegraphics[totalheight=7.0cm]{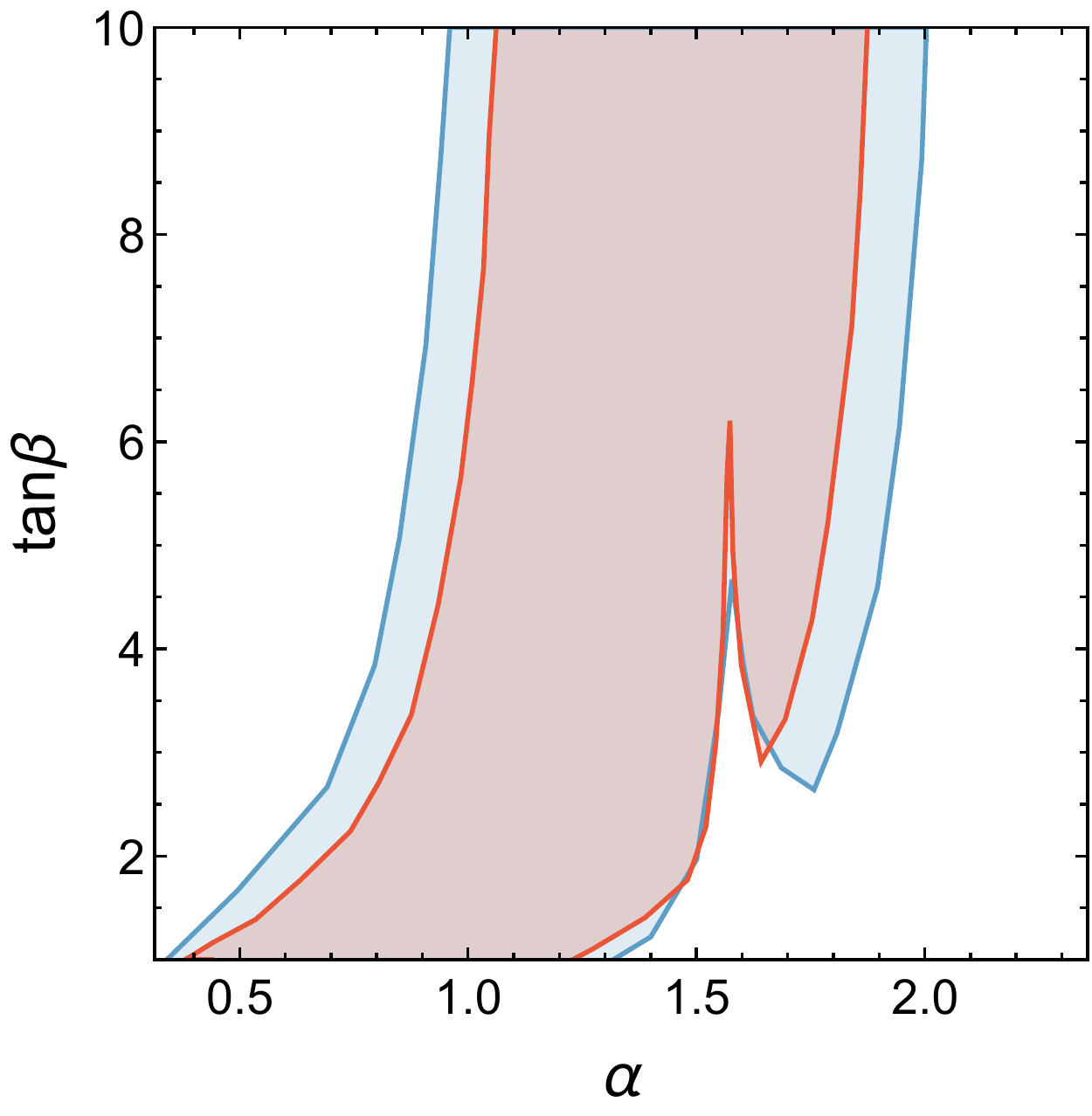}
 \caption{The net allowed parameter space for the type-I 2HDM in the ($\alpha,\tan\beta$) plane
 for $m_h=90$ GeV (pink band) and 100 GeV (blue band) after combing measurements from the Higgs signal strength~\cite{Khachatryan:2016vau} and the direct searches for light scalar at LEP and LHC~\cite{Schael:2006cr,CMS-PAS-HIG-14-037, PhysRevLett.113.171801, CMS-PAS-HIG-17-013}. The wedge-like disallowed region around $\alpha\approx\pi/2$ arises from the direct searches for the light Higgs decaying to the a diphoton at the LHC. This constraint gets relaxed for larger values of $\tan\beta$ and with increasing mass of the light Higgs due to suppression in the production cross section. } 

 \label{fig:LHC}
 \end{center}
 \end{figure} 

Now we combine the individual constraints from the Higgs signal strength measurements~\cite{Khachatryan:2016vau} and the direct searches of the low mass scalars at LEP and LHC~\cite{Schael:2006cr,CMS-PAS-HIG-14-037, PhysRevLett.113.171801, CMS-PAS-HIG-17-013}. The results are shown in Fig.~\ref{fig:LHC} in the ($\alpha,\tan\beta$) plane
for $m_h=90$ GeV (pink band) and 100 GeV (blue band). 
As already noted, the effect of the direct detection constraints from LEP and the Higgs signal strength measurements is to restrict the allowed range of $\alpha$ to be less than $\pi$. In our case, the LEP constraints are far more stringent than those arising from the Higgs signal strength. 
In Fig.~\ref{fig:LHC}, we can see that the allowed range of $\alpha$ increases with as the light Higgs mass increases. This happens due to the relaxed LEP constraints for heavier light Higgs mass (see left panel of Fig.~\ref{fig:sel}). In contrast, the direct search for a light Higgs at the LHC rules out a wedge-like region around the fermiophobic limit and some regions away from the fermiophobic limit. However, the constraints for the latter from the LHC are much weaker than the LEP constraints and consequently are masked in the combination (see Fig.~\ref{fig:LHC}). 
Note that the LHC constraint around the fermiophobic limit gets relaxed for larger values of $\tan\beta$ and with increasing mass of the light Higgs due to suppression in the production cross section.

\section{Future prospects at LHC Run-2}
\label{LHC_future}

In this section, we discuss the prospects of observing a light Higgs boson 
in the following channels: $p p \to h \to \gamma\gamma$, $p p \to W h \to W \gamma\gamma$, $p p \to W h \to b\bar{b}$, and $p p \to t\bar{t} h \to t\bar{t} b\bar{b}$.  
The signal and background processes\footnote{Note that there are two types of backgrounds associated with a particular signal topology: reducible and irreducible.
While the irreducible backgrounds consist of exactly the same final states,
the reducible backgrounds are somewhat different and contribute to a particular signal topology 
because of the misidentification of objects.} corresponding to each channel are generated using the event generators {\tt Madgraph}~\cite{Alwall:2014hca} or {\tt Pythia-8}~\cite{Sjostrand:2007gs, Sjostrand:2014zea}
(depending on the number of final-state hard particles at the parton level)
with the {\tt NN23LO1}~\cite{Ball:2013hta} parton distribution function. The generated events are then showered and hadronized using {\tt Pythia-8}. {Note that the collider analysis has been carried out in {\tt Pythia}. We have not performed any detector simulation in the analysis. 
We now describe the basic cuts used in our analysis.


\begin{enumerate}

\item A minimum cut of 20 GeV is imposed on the transverse momentum of photons, electrons, muons, and missing energy.

\item Owing to the finite resolution of the electomagnetic calorimeter, photons and electrons (muons) are accepted for further analysis if their pseudorapidities are less than $2.5~(2.7)$. 

\item Photons and leptons (electrons and muons) are required to be isolated, 
meaning free from the dominant jet activity in their nearby regions.

\item In experiments, there is a typical $5\%$ probability for an electron to fake a photon, due to track mismeasurements. Since this feature is not present in {\tt Pythia}, we take this into account in our analysis with the help of a random number. We randomly select $5\%$ events, where an electron is mistagged as a photon.

\item  The hadrons are clustered into jets with jet radius $R = 0.4$ using anti-$k_T$ algorithm ~\cite{Cacciari:2008gp}. The jets that satisfy --- $p_{T}^{\rm jet} > $ 30 GeV and $|\eta|<4.5$ are retained for further analysis. 

\item For the topologies which require $b$ tagging, $\Delta R$ is computed between a $b$ parton and each of the anti-$k_T$ jets. If it happens to be less than 0.1, we convolute it with an additional $70\%$ b-tag efficiency factor.

\end{enumerate}

Note that the above cuts (criteria) imposed on the final-state objects in {\tt Pythia} are extremely generic and not specific to any process under consideration. Hence these fall under the category of preselection cuts. 
In the coming sections, we discuss the detailed collider analysis of observing the light Higgs boson. 
The signal significance\footnote{The significance $S$ of observing signal over background 
is defined as $\dfrac{s}{\sqrt{s+b}}$, where $s$ and $b$ are the number of 
signal and background events respectively.} is computed over the allowed parameter space 
as a function of $\alpha$, $\tan\beta$, and $m_h$.

\subsection{Channel 1: $p p \to h \to \gamma\gamma$ }
\label{sec:ggFaa}

We begin with the analysis of the light Higgs boson decaying to the diphoton final state.
For our signal topology, the irreducible background
arises from the tree-level quark-antiquark as well as loop-induced gluon-gluon annihilation to $\gamma \gamma$. The reducible backgrounds 
arise from $j\gamma$, $j j $ and $e^+ e^-$ final states, respectively, where a jet(s) or lepton(s) fakes a photon(s). 
The QCD backgrounds can be considerably reduced by demanding the final-state photons to be isolated (see Table.~\ref{tab:cutflowaa}). The background 
due to the $Z$-pole contributions in the Drell-Yan ($Z \to e e$) process
also dilutes the diphoton signal for light Higgs masses around $m_Z$ due to its large cross section, even though the mistagging rate for an electron to fake a photon is small.

The preselection criteria discussed in the previous section are extremely generic and cannot aid in effective signal-background separation. Additional cuts on the kinematic variables i.e., the transverse momentum ($p_T$) and the invariant mass of the diphoton pair are necessary for further reduction in the background processes. 
To illustrate this point, in Fig.~\ref{fig:pTggF}, we plot the {normalized} transverse momentum distributions for 
the leading isolated photon ($p_{T}^{\gamma}$) corresponding to the signal {(with $m_h$ = 110 GeV)} and the SM backgrounds. 
The $p_T$ for the signal distribution peaks approximately at $m_h/2$ and  
for backgrounds processes (e.g., $\gamma \gamma$, $j \gamma$, and $j j$) it peaks at 
much lower values (although in the plot only 
the $\gamma\gamma$ background is shown). Therefore, suitable choice of the
cuts on the leading and subleading isolated photon candidates and the invariant mass of the diphoton pair can enhance the signal significance. 
The selection cuts used for the diphoton 
analysis are as follows:

\begin{eqnarray}
p_T\,{\rm selection} &:& p_{{ T}_{\rm {lead}}}^\gamma >  40~{\rm GeV} \;, 
p_{{ T}_{\rm {sub}}}^\gamma > 30~{\rm GeV} \;. \\
m^{\gamma\gamma}_{\rm inv}\, {\rm selection} &:&
|m^{\gamma \gamma}_{\rm {inv}} - m_h| < 2.5~{\rm GeV}\;. 
\end{eqnarray}
Here $p_{{ T}_{\text{lead}}}^\gamma$ and $p_{{ T}_{\text{sub}}}^\gamma$ correspond to the transverse momentum of the leading and subleading photon, respectively and $m^{\gamma \gamma}_{\text{inv}}$ corresponds to the invariant mass of the diphoton pair.
Table~\ref{tab:cutflowaa} shows the efficiencies of the preselection and selection cuts on the signal and background processes, where the efficiency of a cut is defined as
\begin{equation}
{\rm Efficiency} \equiv \dfrac{{\rm Number\, of\, events\, after\, imposing\, the\, cut}}{{\rm Number\, of\, events\, before\, imposing\, the\, cut}} 
\end{equation}
\begin{figure} [t]
 \begin{center}
 \includegraphics[totalheight=6.0cm]{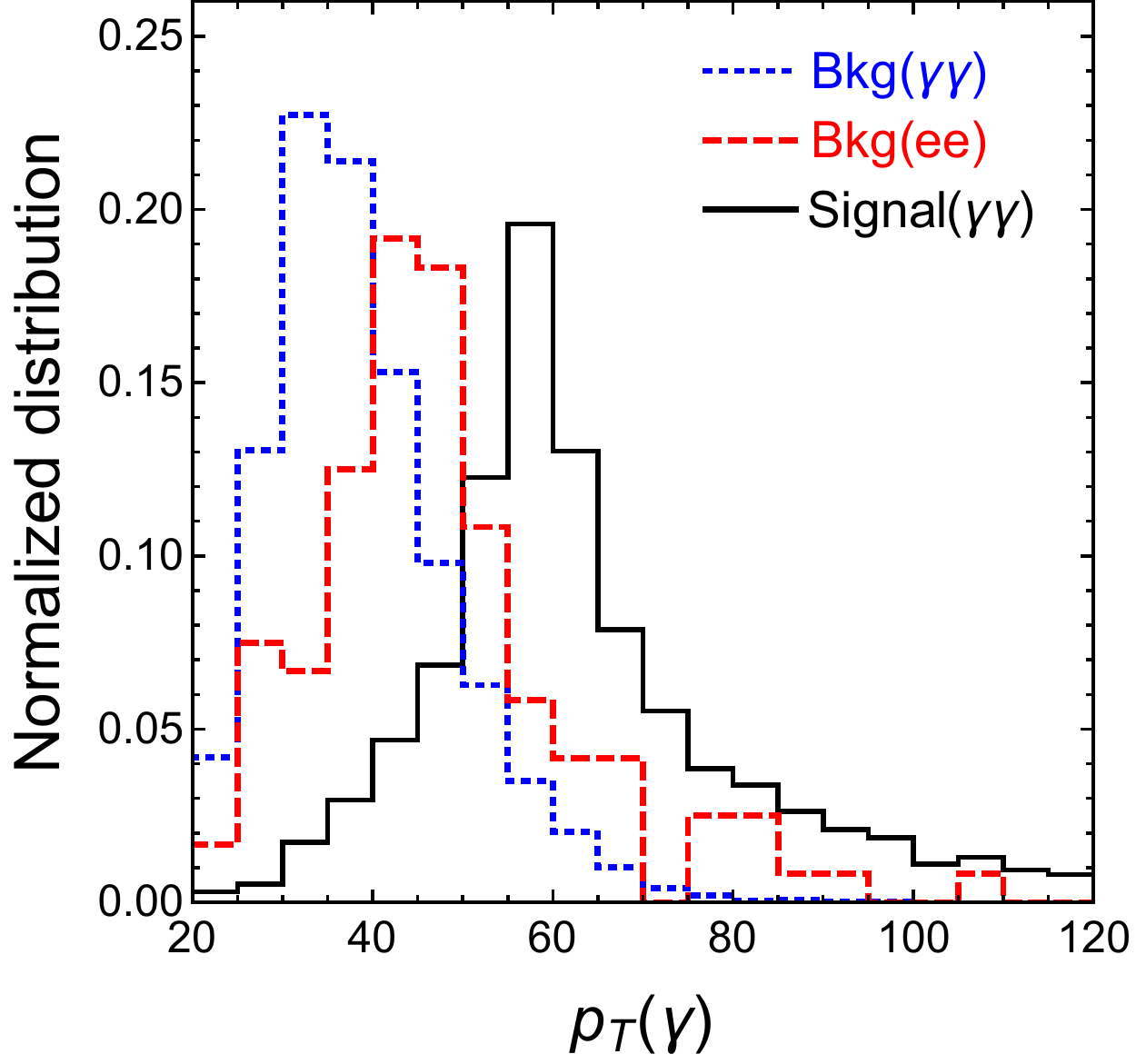}
 \caption{ \textcolor{black}{ The figure illustrates the normalized} $p_T$ distributions of the {leading} isolated photon {in the channel $\gamma \gamma $} for the signal and background processes. 
 \textcolor{black} {Here Signal ($\gamma\gamma$) corresponds to the light Higgs boson of mass $m_h = 110$ GeV, which is produced in the gluon-fusion process and decays to a pair of photons,  Bkg($\gamma\gamma$) corresponds to the irreducible diphoton background and Bkg($ee$) corresponds to reducible background where both electrons fake a photon}.}
 \label{fig:pTggF}
 \end{center}
 \end{figure}

\begin{table}[htb!]
\begin{center}
\def\arraystretch{1.0}
\begin{tabular}{||c|c|c|c|c|c||} \hline \hline
  & \multicolumn{5}{c||}{Efficiency}  \\ \cline{2-6}
 Cuts     & $\;\;\text{Signal}\;\;$ & \multicolumn{4}{c||}{Backgrounds}  \\ \cline{3-6}
        &  &  $\;\;\;\gamma\gamma\;\;\;$        & $\;\;\;j\gamma\;\;\;$ & $j j$ & $ee$  \\ \hline \hline
Preselection & 0.59 & 0.377 & 0.019  & $1.0\times 10^{-6}$ & $1.0\times 10^{-3}$ \\ 
$p_T$ {selection}   & 0.84 &  0.28 & 0.21 & $\sim 0$ & 0.45 \\ 
$m^{\gamma\gamma}_{\rm inv}$ {selection} & 0.99 & 0.082  & 0.024  &  0 & $\sim 10^{-4}$ \\\hline  \hline        
\end{tabular}
\end{center}
\caption{\label{tab:cutflowaa} The efficiencies of the signal and background processes against different cuts are listed for Channel 1. The light Higgs mass is chosen to be $110$ GeV for illustration. The dijet background becomes negligible after imposing all of the cuts.} 
\end{table}


\begin{figure} [t]
\includegraphics[width=0.4\textwidth,height=0.4\textwidth]{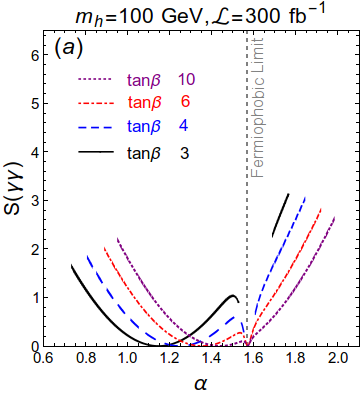}
 \includegraphics[width=0.4\textwidth,height=0.4\textwidth]{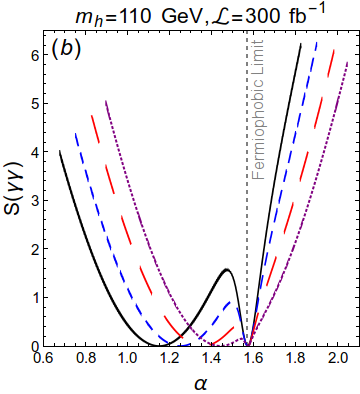}
\caption{ Variation of the signal significance $ S(\gamma\gamma)$ with $\alpha$ for $m_{h} = 100\,,110$ GeV and $\mathcal{L} = 300$ fb$^{-1}$ for different values of $\tan\beta$. Panel (a) corresponds to $m_h=100$ GeV and panel (b) to 110 GeV.
\textcolor{black}{ The vertical gray dashed line corresponds to $\alpha = \pi/2$ i.e., the fermiophobic limit. Here the signal significance drops to zero as expected. Hence, the light Higgs produced in gluon fusion is insensitive to the alignment limit. The discontinuities in panel (a) for $\tan\beta = 3$ and 4 near the fermiophobic limit, correspond to the excluded 
regions from the direct searches of the light Higgs at LHC as discussed in Sec.~\ref{sec:LEP}. }
}
\label{fig:csggF}
\end{figure} 
  
After imposing the preselection and selection cuts on the signal and background processes, we are in a position to determine the signal significance 
for the light Higgs boson as a function of its mass and mixing angles $\alpha$ and $\beta$.
In Fig.~\ref{fig:csggF} (a), we plot the significance of observing $h$ i.e. $S(\gamma\gamma)$ with respect to $\alpha$ for $m_{h} = 100\,$ GeV
and an integrated luminosity $\mathcal{L} =$ 300 fb$^{-1}$ for different values of $\tan\beta$.
In Fig.~\ref{fig:csggF} (b) we repeat the exercise with $m_h=\,$ 110 GeV.  
Note that the significance for smaller masses is negligible, and hence it is not shown in the plot.
We now discuss the qualitative features of Fig.~\ref{fig:csggF} with respect to $\alpha$ and $\tan\beta$.
The discontinuities in Fig.~\ref{fig:csggF} (a) for $\tan\beta = 3$ and 4 near the fermiophobic limit, correspond to the excluded 
regions from the direct searches of the light Higgs at the LHC as discussed in Sec.~\ref{sec:LEP}.
In addition, the constraints from LEP has limited the allowed range of $\alpha$ to be less than $\pi$ for a given light Higgs mass and $\tan\beta$ (see discussions in Sec.~\ref{sec:LEP}).
The dip in the significance signifies the regions 
where the total cross section proportional to ${\xi_h^f} \times {\xi_h^\gamma}$ vanishes. 
The first minimum occurs where $\xi_h^\gamma$ vanishes 
due to cancellation of the top and $W$ loop contribution\footnote{For large values of $\tan \beta$, the dip corresponding to ${\xi_h^\gamma} \to 0$ shifts towards $\alpha \approx \pi/2$ [see Eq.~\ref{(eqn:gammaH)}].} in $h \to \gamma \gamma$ whereas the second minimum corresponds to the fermiophobic limit (${\xi_h^f} \to 0$). \textcolor{black}{Hence, this channel is ineffective in probing the regions close to the fermiophobic limit}.   
The significance of observing the signal
in this channel is larger for $\alpha > \pi/2$ as $\sin(\beta-\alpha)$ is negative in this region. 
As a consequence, the top- and $W$ loop interfere constructively and enhance the 
diphoton rate.

\subsection{Channel 2: $p p \to W h  \to W \gamma\gamma $ }
In this section, we analyze the discovery prospects of the light Higgs boson in the channel $W\gamma\gamma$ at 13 TeV center-of-mass energy,
where the leptonic decays (only $e$ and $\mu$) of $W$ are considered. The SM backgrounds arises from $p p \to W \gamma \gamma$,
$p p \to W j\gamma$, $p p \to W j j$ and $ p p \to W Z\left(Z \to e^+ e^- \right)$. 
The background reduction methods are exactly the same as the ones discussed in Sec.~\ref{sec:ggFaa}, and hence we refrain
from discussing them in this section.

\begin{small}
\begin{table}[h]
\renewcommand{\baselinestretch}{1.0}
\begin{center}
\def\arraystretch{1.2}
\begin{tabular}{||c|c|c|c|c||} \hline \hline
 & \multicolumn{4}{c||}{Efficiency}  \\ \cline{2-5}
  Cuts     & $\;\;\text{Signal}\;\;$  & \multicolumn{3}{c||}{Backgrounds}  \\ \cline{3-5}
        & &  $W\gamma\gamma$        & $\;\;Wj\gamma\;\;$  & $\;\;Wee\;\;$  \\ \hline \hline
{ Preselection} & 0.29 & 0.042 & 0.032 & $4.9 \times 10^{-3}$ \\ \hline 
$p_T$ {selection}    & 0.55 & 0.186 & 0.36 & 0.308 \\ \hline
$m^{\gamma\gamma}_{\rm inv}$ {selection} &   
0.98 & 0.028 & 0.023 & $6\times 10^{-3}$\\ \hline \hline
\end{tabular}
\end{center}
 \caption{\label{tab:cutflowwaa} The efficiencies of the signal and background processes against different cuts for Channel 2. The light Higgs mass is chosen to be $110$ GeV for illustration.} 
\end{table}
\end{small}

The signal is characterized by the presence of at least one isolated lepton, two isolated photons and missing energy. 
The selection cuts used in the analysis are
\begin{eqnarray}
p_T\,{\rm selection}&:&{p}^{\ell}_{ T} > 30~{\rm GeV}\;, {\rm E_T^{miss}} > 30~{\rm GeV}\;, 
p_{{ T}_{\text{lead}}}^\gamma >  40~{\rm GeV} \;,p_{{ T}_{\text{sub}}}^\gamma > 30~{\rm GeV} \;. \nonumber \\
m_{\rm inv}^{\gamma\gamma}\;{\rm selection}&:&|m_{\text{inv}} - m_h| < 2.5~{\rm GeV}\nonumber\;.
\end{eqnarray}
Here ${p}^{\ell}_{T}$ corresponds to the transverse momentum of leptons ($e$ and $\mu$) and 
${\rm E_T^{\text{miss}}}$ denotes the total missing transverse energy. We refer to Table~\ref{tab:cutflowwaa} for the effect of preselection and selection cuts on the signal and background processes.
 This channel allows us to probe the regions close to fermiophobic limit where production via the gluon-fusion process loses its sensitivity. In Fig.~\ref{fig:WHsigmaa}, the significance $S(\ell\nu \gamma\gamma)$ of the signal with respect to $\alpha$ for 100 fb$^{-1}$ integrated luminosity is plotted for four different values of mass of the light Higgs. 
We now summarize the distinctive features of Fig.~\ref{fig:WHsigmaa} below:

\begin{enumerate}
\item For a given light Higgs mass, the significance in this channel decreases as $\tan\beta$ increases, as the production cross section (proportional to $\xi^{V}_{h}$) decreases for large values of $\tan\beta$.  

\item {The branching ratio of $h \to WW^{*}$ increases significantly for larger values of $m_h$. Furthermore the decay, $h \to Z \gamma$ also opens up for $m_h > m_Z$. As a result, the branching ratio of the light Higgs to diphotons decreases with increase in $m_h$. This reduces the signal significance substantially.}

\item The discontinuities in Fig.~\ref{fig:WHsigmaa} correspond to the disallowed regions from the LEP, and LHC direct search measurements.
The chopped-off upper half of the curves in Fig.~\ref{fig:WHsigmaa} (a) for $\tan\beta=4,6$, Fig.~\ref{fig:WHsigmaa} (b) for $\tan\beta=3, 4, 6$ and Fig.~\ref{fig:WHsigmaa} (c) for $\tan\beta =3,4$ near $\alpha \approx \pi/2$ are due to the LHC constraint. These are exactly the disallowed wedge-shaped regions in Fig.~\ref{fig:LHC}. Note that the bounds from the LHC become insignificant for larger values of $\tan \beta$ and $m_h$.
The direct search bounds from LEP on the other hand, constrain the minimum and the maximum values of $\alpha$. This restricts the net allowed range of $\alpha$ to be less than $\pi$. Since in Fig.~\ref{fig:WHsigmaa} we highlight regions close to the fermiophobic limit, the net effect of the LEP constraints is not visible. 

\end{enumerate}     

To conclude, regions around the fermiophobic limit can be best explored at the 13 TeV LHC for lower masses of the light Higgs and intermediate $\tan{\beta}$ values. 

\begin{figure} 
\includegraphics[width=0.4\textwidth,height=0.4\textwidth]{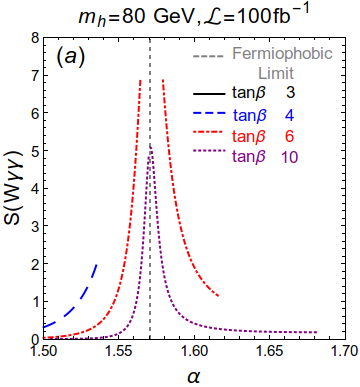}\quad
\quad \includegraphics[width=0.4\textwidth,height=0.4\textwidth]{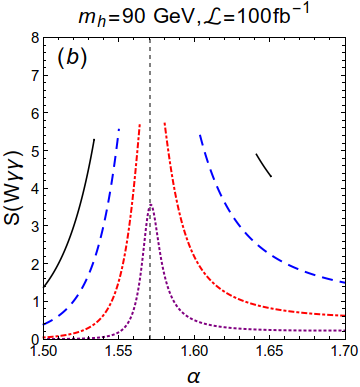}\\
\includegraphics[width=0.4\textwidth,height=0.4\textwidth]{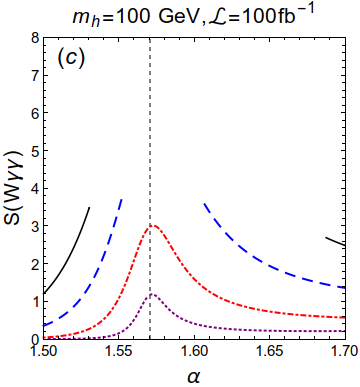}\quad
\quad \includegraphics[width=0.4\textwidth,height=0.4\textwidth]{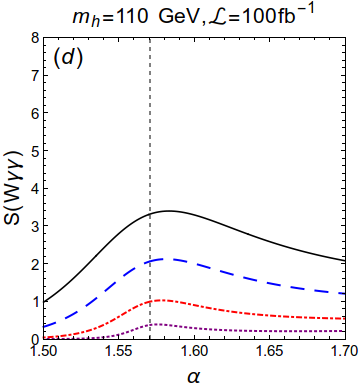}
\caption{ Variation of the signal significance $S(W \gamma \gamma)$ with $\alpha$ for $m_{h} = 80,\,90,\,100,\,110$ GeV for $\mathcal{L}$ = 100 fb$^{-1}$ \textcolor{black}{for different values of $\tan\beta$}. The color code is the same as in Fig.~\ref{fig:csggF}. Note that the range of $\alpha$ is restricted in the plot to signify the regions with reasonable significance. \textcolor{black}{The discontinuities in panel (a) for $\tan\beta = 4$ and $6$, panel (b) for $\tan\beta = 3$, $4$ and $6$, panel (c) for $\tan\beta = 3$ and $4$  near the fermiophobic limit correspond to the excluded regions from the direct searches of the light Higgs at LHC-I as discussed in Sec.~\ref{sec:LEP}. The absence of the $\tan\beta=3$ line in panel (a) is attributed to constraints from LEP which set an upper limit on $\alpha$ and require it to be less than $1.5$ for $\tan\beta=3$ and $m_h=80$ GeV.}}
\label{fig:WHsigmaa}
\end{figure}%

\subsection{Channel 3: $p p \to W h \to W \,b \bar{b} $}

In this section, we analyze the discovery prospects of the light Higgs in the $W b\bar{b}$ channel, where we consider leptonic decays of $W$. 
The signal is characterized by $Wb\bar{b}$, where we tag the leptonic ($e$ and $\mu$) decays of $W$.
The signal is categorized by the presence of two b-tagged jets, an isolated lepton and missing energy. 
In spite of the fact that it is the dominant decay channel in most of the parameter space, the $b\bar{b}$ mode is difficult to probe because of the presence of the enormous QCD background. 
The SM irreducible background arises from $p p \to W Z$. The reducible background arise from $p p \to t\bar{t}$ where one of the $W$'s is along the beam line and hence escapes detection, and $W+$jets where light-quark jets are mistagged as b-jets. The $Wh$ production rate is governed by the magnitude of $\xi^{V}_{h}$ and is small in the favored parts of the parameter space. With a small signal cross section in comparison to large backgrounds, it is difficult to isolate signal events from huge SM backgrounds in the $2b + \ell + E_{T}^{\text{miss}}$ final state at the LHC. In order to achieve appreciable significance at the LHC, we follow the analysis of Ref.\cite{Butterworth:2008iy} and consider the $Wh$ process in the boosted regime. Although we lose a significant number of events by demanding boosted Higgs ($p_{T}^{h} > 200$ GeV), it enables us to overcome huge SM backgrounds quite efficiently. We reconstruct a fat jet with radius parameter $R_{J} = 0.8$ and transverse momentum $p_{T}^{J} > 200$ GeV. We then tag the fat jet as a Higgs using the mass-drop technique discussed in Appendix~\ref{sec:fatjet}. 

The analysis is performed with 14 TeV centre-of-mass energy for $m_{h}$ = 70, 80, 100 and 110. 
We have not considered $m_{h} = 90$ GeV in our analysis as it is   
difficult to isolate the signal from the huge $Z\to b\bar{b}$ background. 
We summarize our selection criteria as follows:
\begin{eqnarray*}
&&p_{T}^{\ell} > 30~\rm{GeV}\;,E_{T}^{\text{miss}} > 30~\rm{GeV} \;, p_{T}^{W} = |{{p_{T}^{\ell}}} + {\bf{p_T^{\text{miss}}}}| > 200\rm{GeV}\;, R_{J} =0.8 \;,\\
&&p_{T}^{J} > 200~\rm{GeV} \;,  |m_{h} - m_{J}| < 5~{\rm GeV} \; \quad \big({\rm for }\; m_h \leq 90~{\rm GeV}\big)\;, \\
&& p_{T}^{J} > 250~{\rm{GeV}}  \;,|m_h - m_{J}| < 8~{\rm GeV}
\; \quad \big({\rm for }\; m_h > ~90~{\rm GeV}\big)\;,
\end{eqnarray*}
where ${p_{T}^{W}}$ is the magnitude of the vector sum of the momentum of the lepton and missing energy in the transverse plane.
{The efficiencies of these cuts are displayed in Table.~\ref{tab:cutflowwbb}. We can see that by demanding at least one fat jet and anti-$k_T$ jet reduces $Wb\bar{b}$ and $W3j$ backgrounds. Also, by invoking a fat jet with no jet activity outside and MassDrop with a double b-tag, we are able to suppress the $t\bar{t}b\bar{b}$ process very effectively.}

\begin{figure}
\includegraphics[width=0.4\textwidth,height=0.4\textwidth]{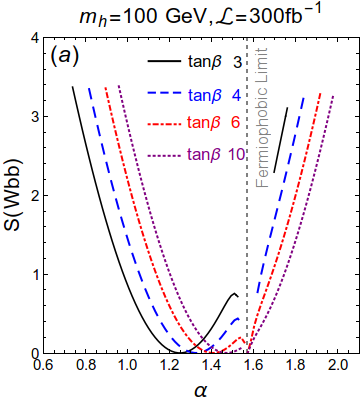}
\quad\includegraphics[width=0.4\textwidth,height=0.4\textwidth]{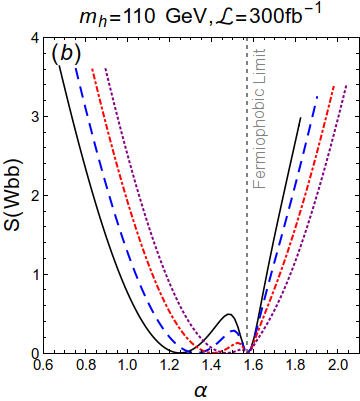} 
  \caption{Variation of the signal significance $S(W bb)$ with $\alpha$ for $m_{h} = 80,\,90,\,100,\,110$ GeV for $\mathcal{L}$ = 300 fb$^{-1}$ \textcolor{black}{for different values of $\tan\beta$}. The color code is the same as in Figs.~\ref{fig:csggF} \ref{fig:WHsigmaa}. This channel is also insensitive around the fermiophobic limit.
 }
 \label{fig:whbba}
\end{figure}
After imposing the above cuts, we compute the signal significance 
for the light Higgs boson as a function of its mass and mixing angles $\alpha$ and $\beta$. In Fig.~\ref{fig:whbba} we plot the significance  
of observing the light Higgs as a function of $\alpha$. 
Figures~\ref{fig:whbba} (a) and \ref{fig:whbba} (b) represent the significance with an integrated luminosity of 300 fb$^{-1}$ for $m_{h} = 100$ and 
$110$ GeV respectively.  Again, the discontinuities in Fig.~\ref{fig:whbba} arise due to direct detection constraints from LEP and LHC.   
It is interesting to note the behavior of the signal significance in Fig.~\ref{fig:whbba}. The dip in the plot signifies the points where the total cross section proportional to $\xi^{V}_{h} \times \xi^{f}_{h}$ vanishes. The first dip corresponds to $\xi^{V}_{h}\to 0$ and the second dip represents $\xi^{f}_{h}\to 0$ (fermiophobic limit).
Hence, this channel is useful in probing regions away from the fermiophobic limit.

\begin{small}
\begin{table}[h]
\begin{center}
\def\arraystretch{1.0}
\begin{tabular}{||c|c|c|c|c||} \hline \hline
& \multicolumn{4}{c||}{Efficiency}  \\ \cline{2-5}
  Cuts      & $\;\;\rm{Signal}\;\;$ & \multicolumn{3}{c||}{Backgrounds}  \\ \cline{3-5}
       &  &  $\;\;Wbb\;\;$ & $\;\;W3j\;\;$ & $\;\;t\bar{t}b\bar{b}\;\;$ \\ \hline \hline
 At least one fat jet and anti-$k_T$ jet & 0.45 & 0.11 & 0.10 & 0.47 \\ 
 Isolated leptons & 0.86 & 0.71 & 0.68 & 0.21 \\ 
 One fat jet with no anti-$k_T$ jet & 0.5 & 0.27 & 0.16 & 0.019 \\ 
 $E_T^{{\rm miss}} > 30$ GeV & 0.987 & 0.93 & 0.93 & 0.99 \\ 
 $p_{T}^{W} > $ 200 GeV & 0.93 & 0.88 & 0.85 & 0.77 \\
 MassDrop with double b-tag & 0.32 & 0.299 & 0.0037 & 0.031 \\
 Inv. mass & 0.79 & 0.077 & 0.077 & 0.11 \\ \hline   \hline
\end{tabular}
\end{center}
 \caption{\label{tab:cutflowwbb} The efficiencies of the different cuts used for the analysis of Channel 3 for both signal and background processes. The numbers are for a light Higgs mass of 110 GeV.}
\end{table}
\end{small}

\subsection{Channel 4: $p p \to t \overline{t} \,h  \to t \overline{t} \,b \overline{b} $ }
Continuing with the discussion of a light Higgs decaying to $b\bar{b}$, we now 
focus our attention on the $t \bar{t} h$ production mode, where semileptonic decays of top-pair are considered
The irreducible background here arises from the $t\bar{t}b\bar{b}$ final state and the reducible background arises from $t\bar{t} + \rm{jets}$, where a jet fakes the bottom quark. Due to the presence of four b quarks in the final state, it is difficult to reconstruct the light Higgs accurately due to the various possible combinations. 
This problem can be addressed by resorting to boosted scenarios where 
%
\begin{figure}[t]
\includegraphics[width=0.4\textwidth,height=0.4\textwidth]{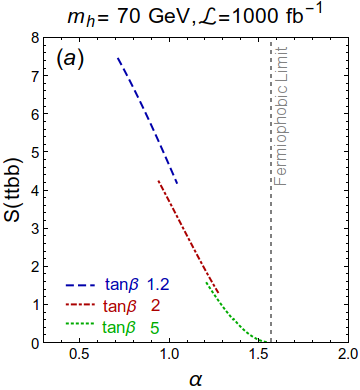}\quad
\includegraphics[width=0.4\textwidth,height=0.4\textwidth]{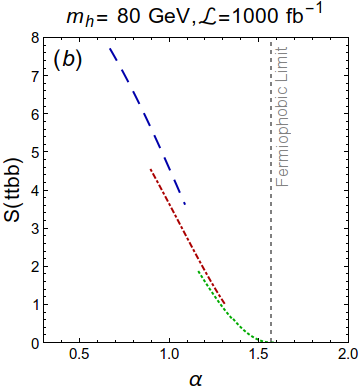}\\ 
\quad\includegraphics[width=0.4\textwidth,height=0.4\textwidth]{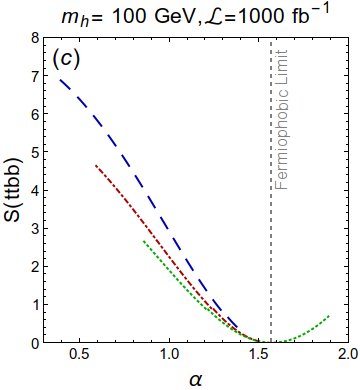}
\quad\includegraphics[width=0.4\textwidth,height=0.4\textwidth]{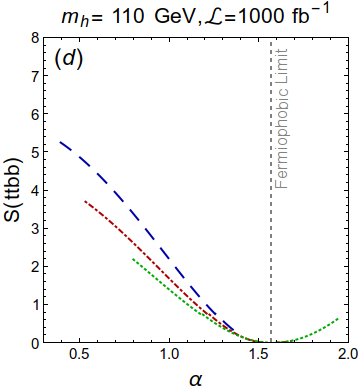}
\caption{Variation of the signal significance $S(t \bar{t} b \bar{b})$ in the channel  $t\bar{t}b\bar{b}$  with $\alpha$ for different values of $m_{h}$ at 1000 fb$^{-1}$ integrated luminosity. While the dark blue, dark red, and dark green dashed lines correspond to $\tan\beta =$ 1.2, 2, and 5, respectively, the dashed gray vertical line for $\alpha=\pi/2$ illustrates the fermiophobic limit. Owing to enhanced sensitivity of $\sigma(pp\to t\bar{t}h \to t\bar{t}b\bar{b})$ for low $\tan\beta$, we have chosen slightly lower values of $\tan\beta$ for this channel.}
 \label{fig:tthbba}
\end{figure}
the decay products of the hadronically decaying top and light Higgs are enclosed within a single jet of large radius parameter. Therefore, our signal essentially comprises of two fat jets, an isolated lepton, missing energy and one anti-kT b tagged jet. To tag the top and Higgs jets, we first construct the fat jets with $p_{T}^{J} > 125$ GeV and $\Delta R = 1.2$. The jets satisfying $p_{T}^{J} > 250$ GeV are tagged as top-jet if they satisfy the prescription described in appendix.~\ref{sec:fatjet}. Similarly the remaining jets are tagged as the Higgs jets if they satisfy the mass drop criteria and the filtered jet mass, $m_J^{\text{Higgs}}$, lies within 5/10 GeV window about the light Higgs mass (see appendix.~\ref{sec:fatjet} for more details). In addition, we demand a b-tagged jet outside the top and Higgs fat jet. This helps in further eliminating the $t\bar{t}+$jets background. We summarize the cuts used in the analysis below:
\begin{eqnarray*}
&& p_{T}^{\ell} > 30~\rm{GeV}\;,E_{T}^{miss} > 30~\rm{GeV} \;,
p_{T}^{\text{top}} > 250~{\text{GeV}},~150~{\rm{GeV}} < m^{\text{top}}_{J} < 200~{\rm{GeV}}\;,\\
&& p_{T}^{\rm Higgs} > 125~{\rm{GeV}}  \;,|m_J^{\text{Higgs}} - m_{h}| < 5~{\rm GeV} \;\; \quad \big({\rm for }\; m_h \leq 90{\rm GeV}\big) \\
&& p_{T}^{\rm Higgs} > 160~{\rm{GeV}}  \;,|m_J^{\text{Higgs}} - m_{h}| < 10~{\rm GeV}
\; \quad \big({\rm for }\; m_h >  90~{\rm GeV}\big)\;.
 \end{eqnarray*}
{The efficiencies of the individual cuts are listed in Table~\ref{tab:cutflowttbb}.}
We are now in a position to estimate the signal significance i.e., $S(t \bar{t} b \bar{b})$ as a function of $\alpha$, $\tan\beta$, and $m_h$. In Fig.~\ref{fig:tthbba} we plot the significance of observing a light Higgs for four different light Higgs masses: $m_h = $ 70, 80, 100, and 110 GeV. 
We have not considered $m_h=$ 90 GeV for the analysis because in that case it will be difficult to isolate the signal events from the large $t\bar{t}Z$ background.
Note that we have chosen smaller $\tan\beta$ values as the total cross section decreases with increase in $\beta$ (see Table.~\ref{tab:crosssection-behaviour}). The significance is higher for lower values of $\alpha$. Hence this channel is effective for probing lower $\tan{\beta}$ and $\alpha$ regions. This particular mode for probing the light Higgs does not work out in the fermiophobic limit as both the production cross section and decay branching ratio are negligible.

\begin{small}
\begin{table}[h]
\begin{center}
\def\arraystretch{1.0}
\begin{tabular}{||c|c|c|c||} \hline \hline
 & \multicolumn{3}{c||}{Efficiency}  \\ \cline{2-4}
 Cuts     & $\;\;\text{Signal}\;\;$ & \multicolumn{2}{c||}{Backgrounds}  \\ \cline{3-4}
        &  &  $\;\;ttbb\;\;$ & $\;tt+3j\;$ \\ \hline \hline
 Isolated leptons & 0.53 & 0.56 & 0.57  \\ 
 Two fat jets & 0.31 & 0.17 & 0.20 \\ 
 $p_T^{{\ell}} > $ 30 GeV, $E_T^{{\rm miss}} > 30$ GeV & 0.76 & 0.65 & 0.63 \\ 
 Top tagged & 0.11 & 0.088 &  0.13 \\
 Mass drop with double b-tag and inv. mass   & 0.056 & 0.011 & 0.0009  \\
 Anti-$k_T$ b-jet outside top and Higgs jet & 0.28 & 0.25 & 0.50  \\ \hline \hline 
\end{tabular}
\end{center}
 \caption{\label{tab:cutflowttbb} The efficiencies of the different cuts used for the analysis of Channel 4 for both signal and background processes. The numbers are for a light Higgs mass of 110 GeV.}
\end{table}
\end{small}


\section{Summary and concluding remarks}
\label{sec:discuss}

To summarize, we studied the prospects of observing a CP-even scalar lighter than the observed 125 GeV Higgs at the LHC, in the context of the type-I 2HDM. We identified the heavier CP-even Higgs in the 2HDM with the discovered Higgs. We also considered 
the charged and pseudoscalar Higgs bosons to be heavy. This choice simplifies the 2HDM parameter space and leaves $\alpha$, $\tan\beta$, and the mass of the light Higgs ($m_h$) as the relevant free parameters. We considered various theoretical and experimental constraints to determine the allowed regions in the parameter space. 
The mass of the light Higgs was taken to be greater than 62.5 GeV to avoid $H \to h h$ decay. 

To study the phenomenology of the light Higgs at the LHC, we determined the suitable production and decay modes. In most parts of the parameter space, the light Higgs in the type-I 2HDM decays dominantly to $b\bar{b}$. However, for regions close to the fermiophobic limit, its decay to bosons (mainly photons) becomes dominant. Therefore, we focused on the light Higgs decay to $b\bar{b}$ and $\gamma\gamma$ in this analysis.  
Analyzing $b\bar{b}$ in the $ggF$ or VBF production mode is challenging due 
to the large QCD background. We chose the light Higgs production in association with the $W$ boson and top pair for the $b\bar{b}$ analysis. Furthermore, we tagged the light Higgs in the boosted regimes, for better signal significances. The choice of the production mode for the $\gamma\gamma$ channel is much simpler because of its better reconstruction properties. We chose the $Wh$ production mode to analyze regions close to the fermiophobic limit, and the $ggF$ production mode for regions away from the fermiophobic limit. 

\begin{figure}[t]
\begin{center}
\includegraphics[width=0.7\textwidth]{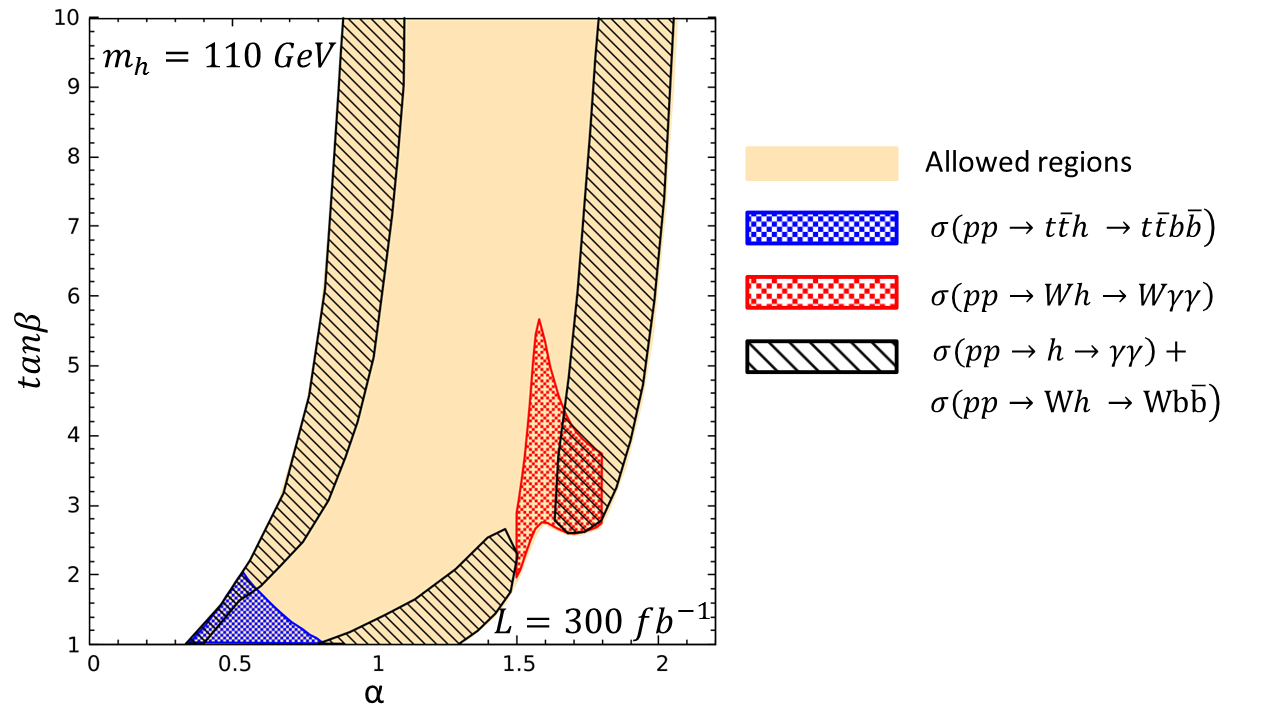}
\caption{\textcolor{black}{\label{fig:summary} Regions of the allowed parameter space that could be probed/excluded in different channels with significances greater than $2\sigma$ for a light Higgs boson of mass $110$ GeV at the LHC with $300$ fb$^{-1}$ of integrated luminosity. While the  yellow contour illustrates the total allowed region in the $\left(\alpha,\tan\beta\right)$ plane for $m_h = 110$ GeV, we can probe only the hatched regions with significances greater than $2\sigma$, leaving behind regions which satisfy the alignment limit $\alpha\approx \beta$. Here we have combined the allowed regions for $pp\to h \to \gamma \gamma$ and $pp\to W h \to W b\bar{b}$ due to their similar behavior.}}
 \end{center}
\end{figure}

We analyzed the discovery prospects of the light Higgs boson in four channels atthe LHC: $p p \to h \to \gamma\gamma $, $p p \to W h  \to W \gamma\gamma $, $p p \to W h \to W b \bar{b} $, and $p p \to t\bar{t} h \to t \bar{t} b \bar{b}$ at the LHC. 
We found interesting regions in the parameter space of the 2HDM that could be probed at the future runs of the LHC with a few hundred fb$^{-1}$ of luminosity. We summarize our findings in 
Fig.~\ref{fig:summary} for $m_h=110$ GeV and $\mathcal{L} = 300$ fb$^{-1}$.  In this plot, the yellow contour illustrates the total allowed region for $m_h=110$ GeV. The hatched portions denote the regions where the above channels could be probed with significances greater than 2$\sigma$ at the LHC. The unhatched regions in the allowed contour correspond to
$\alpha \approx \beta$ and approximately satisfy the alignment limit. 
As already noted, such regions would be difficult to probe/rule out in the near future.
For the purpose of the plot, we have combined the allowed regions for $pp\to h+X \to \gamma \gamma+X$ and $pp\to W h + X \to W b\bar{b} + X$ as they probe almost similar parts of the parameter space (see Appendix~\ref{sec:csplot}).

Searches for physics beyond the Standard Model to date have yielded neither any significant results nor specific directions to follow. 
However, the current measurements still do not rule out the possibility of the observed 125 GeV scalar belonging to some enlarged sector. 
In this paper, we examined a possible scenario in context of the type-I 2HDM and studied the prospects of observing a light CP-even scalar at the future runs of the LHC.
Our aim in this study was to put together all of the relevant 
information and provide an optimized search strategy for the light Higgs at the LHC.   
The discovery of such a light scalar would not only open doors to new physics but also help us to better understand the mechanism of electroweak symmetry breaking.

\section{Acknowlegements}
We would like to thank Sandhya Jain, Manoranjan Guchait, and Tuhin Roy for helpful discussions. We would also like to acknowledge Amol Dighe, Debjyoti Bardhan, Sreerup Raychaudhuri, and Tuhin Roy for careful reading of our manuscript. S.N. acknowledges the Dr. D. S. Kothari Post Doctoral Fellowship awarded by University Grant Commission (award letter no. PH/15-16/0073) for financial support. We would also like to thank the Department of Theoretical Physics, TIFR for the use of its computational resources.

\noindent{\textbf{\large{Note added:}} Recently the CMS collaboration reported a local excess of $2.8 \sigma$ (global 1.3 $\sigma$)in the $\gamma\gamma$ channel around 95 GeV~\cite{CMS-PAS-HIG-17-013}. Although at this stage the deviation is not significant, but is certainly of importance simply because LEP had also indicated a possibility of
observing an excess of Higgs-like events in similar mass region~\cite{Barate:2003sz}. Therefore, it is important to analyze theories that predict low mass scalar and which can explain such excess. Type-I 2HDM, in this context, perfectly fits the bill. Since the excess at present is only indicative, we have restricted ourselves to a generic low mass analysis in the type-I 2HDM. In particular, for implications of the type-I 2HDM in the light of the recent CMS result, one could refer to \cite{Fox:2017uwr,Haisch:2017gql}.}


\appendix

\section{Diphoton loop}
\label{sec:loop}
The effective interactions of $h(H)$ with $\gamma \gamma$ are given as~\cite{Gunion:425736}:
\begin{eqnarray}
L &=& \frac{\alpha_{em}}{8 \pi v}\; \xi^{\gamma}_{h} h\; F_{\mu \nu} F^{  \mu \nu}  
 \;\;+\frac{\alpha_{em}}{8 \pi v}\; \xi^{\gamma}_{H} \;H F_{\mu \nu} F^{ \mu \nu} \;.
\end{eqnarray} 
Correspondingly the decay width is
\begin{eqnarray}
\Gamma(h(H) \to \gamma \gamma) = \frac{\alpha^2 g^2}{1024 \pi^3} \frac{m_{h(H)}^3}{m_W^2} |\xi^{\gamma}_{h(H)}|^2\;.
\end{eqnarray} 
For the type-I 2HDM, the effective couplings $\xi^{\gamma}_{h(H)}$ receive dominant contributions from $W$-boson, charged Higgs, and top loop, and are given as
\begin{eqnarray}
 \xi^{\gamma}_{h(H)} &=& N_c Q_t^2 \;\xi^{t}_{h(H)}  F_{1/2} (\tau_{t}) 
                   + \xi^{W}_{h(H)} F_{1} (\tau_{W}) 
                   + \frac{m_W^2}{M^2_{H^\pm}}\xi^{H^{\pm}}_{h(H)} F_{0}(\tau_{H^{\pm}}) \;.
\label{eqn:gammaH}                   
\end{eqnarray}
 The form factors are given as
 \begin{eqnarray}                  
  F_{0}(\tau_{H^\pm}) &=& \tau_{H^\pm} \left[ 1 - \tau_{H^\pm} f (\tau_{H^\pm}) \right] \;, \quad
F_{1/2}(\tau_{t}) = -2\tau_t \left[ 1 +  (1 - \tau_t) f(\tau_t) \right]\;,
\\
&&F_{1}(\tau_{W}) = 2 + 3 \tau_W + 3 \tau_W (2 - \tau_W) f(\tau_W) \;,
\end{eqnarray}
where 
\begin{eqnarray}
f(\tau) &=&  \left(\sin^{-1}\frac{1}{\sqrt{\tau}}\right)^2~ {\rm{for}}~ \tau > 1\;, \quad\quad f(\tau) = -\frac{1}{4}\left(\log \frac{\eta_+}{\eta_-}-i \pi\right)^2~ {\rm{for}}~ \tau < 1 \;, \nonumber\\
\eta_\pm &=& 1\pm \sqrt{1- \tau}\;, \quad \quad \quad \quad \quad \quad \quad\quad \quad \tau = 4 \left(m/m_{h(H)} \right)^2  \;.
\label{loop}
\end{eqnarray}
The couplings of $h(H)$ with $t\bar{t}$, $W^+W^-$, and $H^+ H^-$ in the type-I 2HDM are
\begin{eqnarray}
\xi^{t}_{h} &=& \cos{\alpha}/\sin{\beta}\;, \xi^{t}_{H} = \sin\alpha/\sin\beta\;,
\xi^{W}_{h} = \sin{(\beta-\alpha)}\;, \xi^{W}_{H} = \cos{(\beta-\alpha)}\;,\\
\xi^{H^\pm}_{h} &=& \dfrac{1}{4 m_W^2 \sin^2(2\beta)}\bigg[ 8 m_{12}^2 \cos{(\alpha+\beta)} - 
\sin(2\beta) \bigg( (m_h^2 - 2 m_{H^\pm}^2) \cos(\alpha-3\beta) \nonumber \\
&& \quad \quad \quad \quad+ (2 m_{H^\pm}^2 + 3 m_h^2) \cos(\alpha+\beta) \bigg) \bigg] \;,\\
\xi^{H^\pm}_{H} &=& \dfrac{1}{4 m_W^2 \sin(2\beta)}\bigg[ ( 2 m^2_{H^\pm} - m_H^2)  \sin{(\alpha-3 \beta)} + 
\sin(\alpha+\beta) \bigg( \dfrac{4 m_{12}^2}{\sin\beta \cos\beta} \nonumber \\
&& \quad \quad \quad \quad - 2 m^2_{H^\pm} - 3 m_H^2 \bigg) \bigg]\;.
\end{eqnarray}

\section{Charged Higgs analysis}
\label{sec:chargedHiggs}

In this appendix, we revisit some of our analyses by considering effect of a low-mass charged Higgs. We will see that our results will remain more or less unaltered.
The independent 2HDM parameters are varied in the following ranges{\footnote{The lower range of $M_{A}$ has been kept the same as that of $M_{H^{\pm}}$ for simplicity.}}:
\begin{eqnarray}
\alpha = [0,\pi]\;, \tan\beta=[1,10]\;, m_{12}=[0.01,1000] \;{\rm GeV}\;, \nonumber \\
M_A =[80,2000] \; {\rm GeV} \;, M_{H^{\pm}} = [80,2000]\; {\rm GeV}\;.
\end{eqnarray}

As in the tType-I 2HDM, couplings decrease as $\tan\beta$ increases, and we have fixed the upper limit on $\tan\beta$ to be 10. We first determine the allowed parameter space by incorporating the following constraints:

\begin{itemize}
\item Perturbativity: We demand that the 
Higgs self couplings, i.e., $\lambda_{i}$ and the Yukawa couplings be less than $4\pi$
for the perturbative expansion to remain valid. 

\item Vacuum stability: This condition ensures that the scalar potential is bounded from below by restricting the $\lambda_i$'s in the ranges \newline
$\lambda_{1,2} > 0$, $\lambda_3 > -\sqrt{\lambda_1 \lambda_2}$
and $\lambda_3 + \lambda_4 - |\lambda_5| > -\sqrt{\lambda_1 \lambda_2}$ ~\cite{Nie:1998yn}.
\item Unitarity: This arises from the requirement of unitarity of the scattering amplitudes such that the amplitudes do not grow as the center-of-mass energy increases. The unitary bounds for the 2HDM can be found in Ref.~\cite{Akeroyd:2000wc}
\item $\rho$-parameter $\left(\dfrac{m_W^2}{m_Z^2\cos \theta_W}\right)$: Its value in the SM is predicted to be unity at tree level (the renormalization scheme is chosen such that this relation even holds after including higher-order corrections~\cite{Ross:1973fp}). The experimental prediction of $\rho$ parameter is in agreement with the SM and constrains the masses of new scalars introduced in the theory.~\cite{ALEPH:2005ab}.

\item Flavor observables:~
Although the tree-level FCNCs in the 2HDM are absent due to the $Z_{2}$ symmetry, the charged scalars can affect these processes through higher-order diagrams. In general, the flavor observables in these models are sensitive to $m_{H^{\pm}}$ and $\tan \beta$.

\item Direct charged Higgs searches at LEP: 
The charged Higgs has been searched for in the channel $e^{+} e^{-} \to H^{+}H^{-}$ at LEP. The null observation of the signal has put 
 a lower bound of 80 GeV on the mass of charged Higgs~\cite{Searches:2001ac,abreu:2001ib}. This bound has been derived assuming $H^\pm$ decays only to the $\tau\bar{\nu}$ and $c\bar{s}$ modes. However, in the alignment limit the decay $H^\pm \to h W^\pm$ becomes significant. Hence, the bound on the charged Higgs mass gets relaxed in the regions close to the alignment limit~\cite{Arbey:2017gmh}. 

\end{itemize}
 
The effect of the above constraints on the parameter space is shown in Fig.~{\ref{fig:chargedHiggs}}. The allowed regions are shown in Fig.~{\ref{fig:chargedHiggs}}(a), Fig.~{\ref{fig:chargedHiggs}}(b), and Fig.~{\ref{fig:chargedHiggs}}(c) of in the ($M_{A},M_{H^\pm}$), ($\alpha,m_{12}$), and ($M_{H^\pm},m_{12}$) planes, respectively.
For the type-I model, the bounds from flavor physics are weak and allow almost all values of $m_{H^{\pm}} \gtrsim 80$ GeV for $\tan \beta \gtrsim 2$~\cite{Mahmoudi:2009zx,Aoki:2009ha,Arbey:2011zz}. Hence we have not shown the effect in the plot.

The important inferences which we can make from Fig.~\ref{fig:chargedHiggs} are as follows:
\begin{enumerate}
\item It can be seen from Fig.~{\ref{fig:chargedHiggs}} (a) that there exist upper bounds on the masses of the charged Higgs and the pseudoscalar Higgs. These bounds arise primarily due to the unitarity constraints. 
Furthermore bounds from the $\rho$ parameter force the mass of the pseudoscalar to be approximately equal to that of the charged Higgs for $m_{H^\pm} \gtrsim 200$ GeV, and for $m_{H^\pm} \lesssim 200$ GeV, the pseudoscalar
mass remains unconstrained. 

\item The $Z_2$
symmetry-breaking parameter $m_{12}$ is also restricted to be less than 100 GeV [see Figs.~\ref{fig:chargedHiggs} (b) and \ref{fig:chargedHiggs}(c)]. These bounds arise from the vacuum stability requirements.

\item The mixing angle $\alpha$ is not constrained at all by any of the above constraints, as can be seen in Fig.~{\ref{fig:chargedHiggs}} (c).

\end{enumerate}

\begin{figure}[h] 
\begin{center}
\includegraphics[height=4.0cm,width=0.31\textwidth]{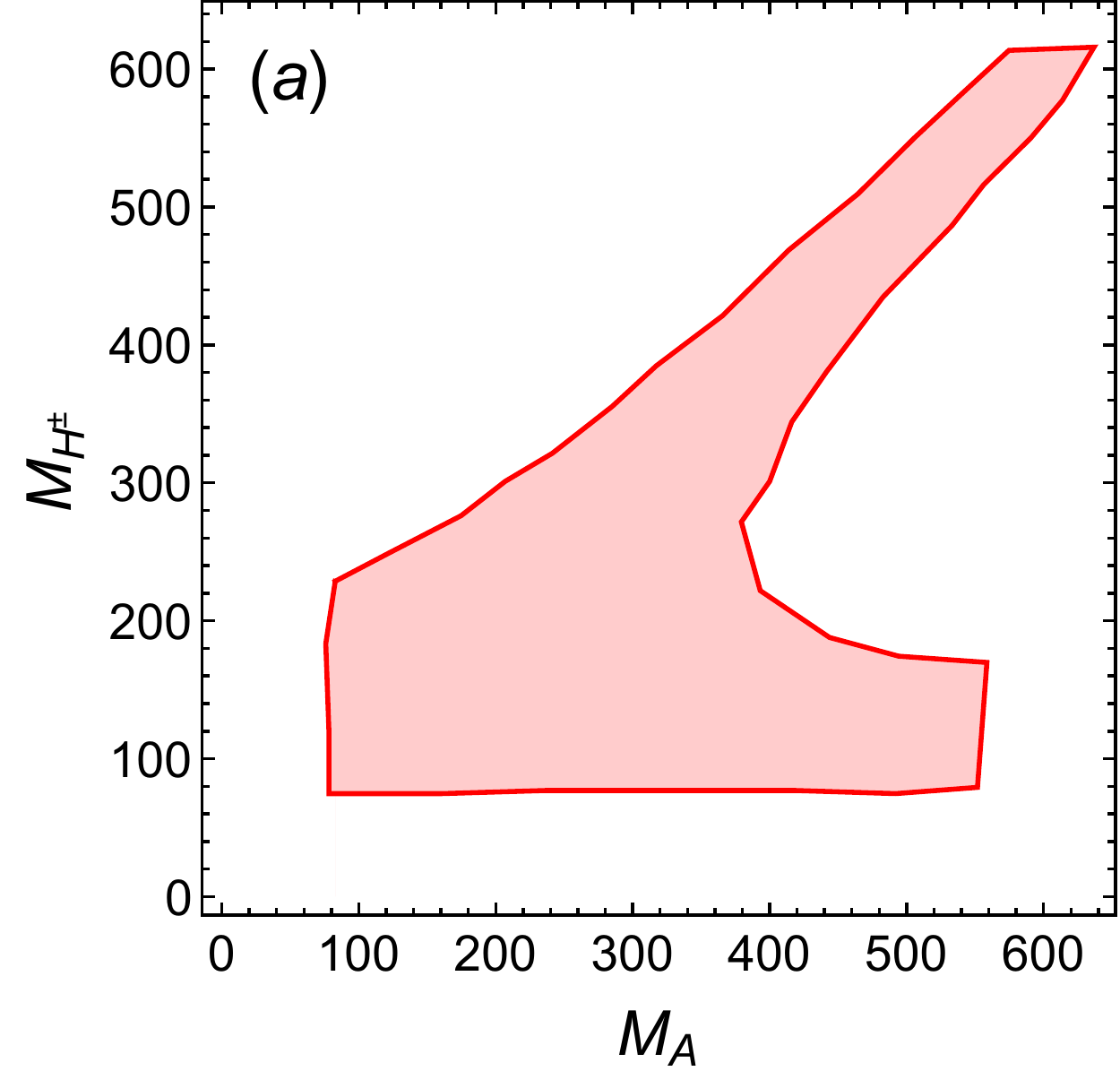}
\includegraphics[height=4.0cm,width=0.31\textwidth]{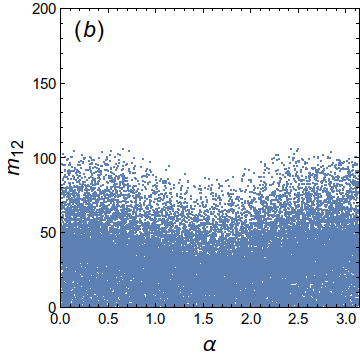}
\includegraphics[height=4.0cm,width=0.31\textwidth]{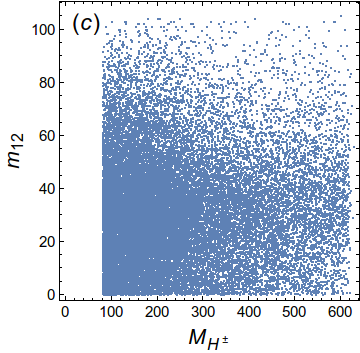}
\caption{ Allowed range of $M_A$ and $M_{H^\pm}$ after imposing constraints from perturbativity, vacuum stability, tree-level unitarity, $\rho$-parameter, LEP and flavour data.
}
\label{fig:chargedHiggs}
\end{center}
\end{figure} 
%
%
After determining the allowed parameter space from theoretical and a few experimental constraints, we proceed to examine the effect of a light charged Higgs on the allowed parameter space from the Higgs signal strength 
measurements (for earlier analyses of this kind, see Ref.~\cite{Posch:2010hx}). For illustrative purposes, we have fixed the mass of the charged Higgs to be 200 GeV
and $m_{12}^2 = 100$ GeV. 
The charged Higgs boson will affect the signal strength measurements through its contribution in $H\to\gamma\gamma$ decay. 
It can be seen from Fig~\ref{fig:chargedHiggs2} that 
the deviations in the high \-$\tan\beta$ regions are dramatic while for low-$\tan\beta$ the increment in the allowed range of $\sin(\beta-\alpha)$ is slight. Furthermore, for the low-$\tan \beta$ regions the LEP measurements are far more constraining (see Fig.~\ref{fig:LHC}). 
Therefore, the allowed parameter space for $\tan\beta < 10$ (which is our region of interest) remains the same even after including effects from the low-mass charged Higgs.

A light charged Higgs boson could also affect the significance of observing a light Higgs $h$ in the $\gamma \gamma$ channel. However, it is found that the significance only increases slightly for larger values of $|\sin(\beta-\alpha)|$. The effect is depicted in the right panel of Fig~\ref{fig:chargedHiggs2}. Although the plot is shown for particular choices of $m_h$ and $\tan\beta$, the qualitative result is independent of their values.
Hence, the effect of considering a light charged Higgs boson only mildly affects our analyses. 

\begin{figure}[h] 
\begin{center}
\includegraphics[width=0.35\textwidth]{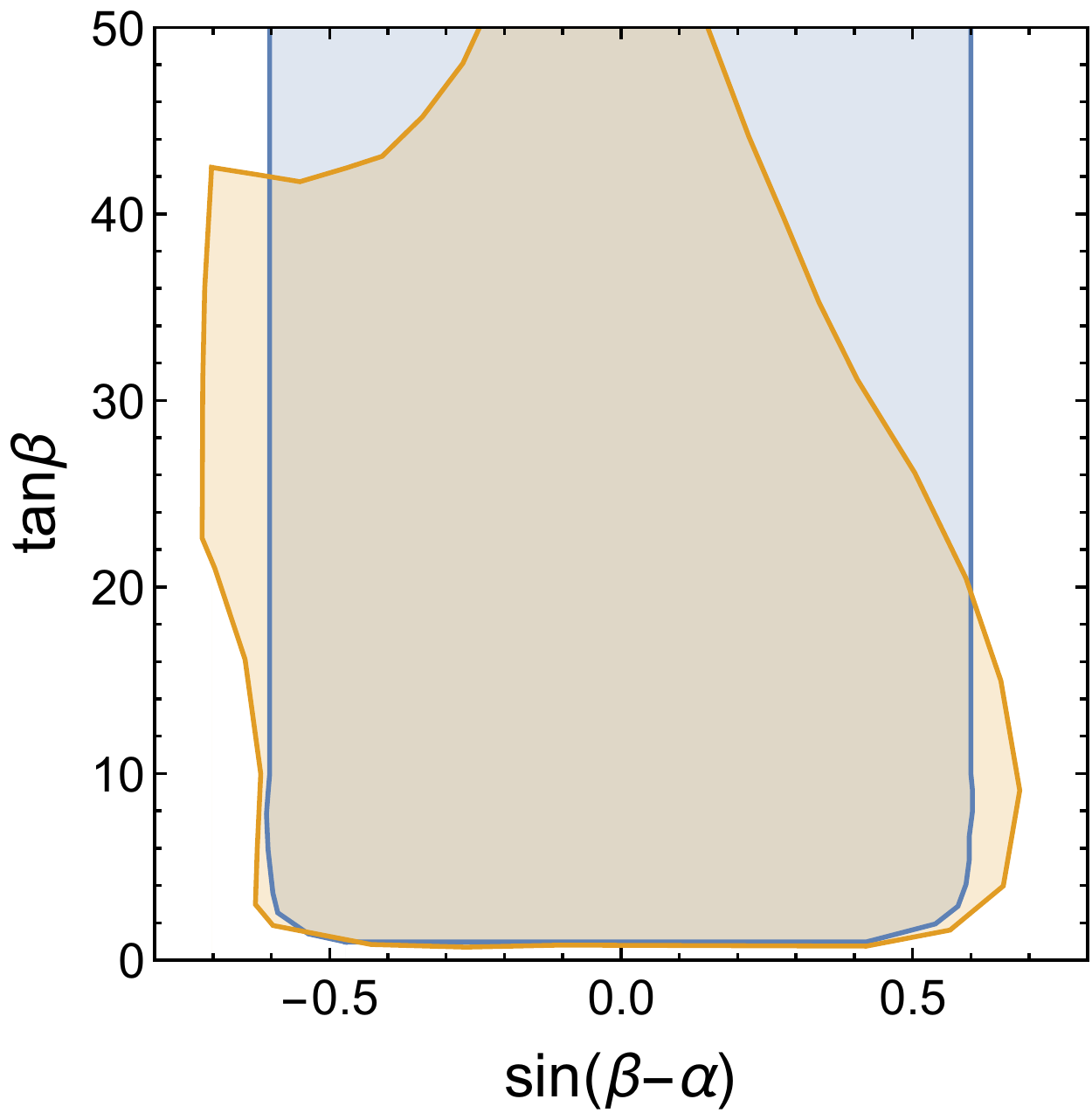}
\includegraphics[width=0.35\textwidth]{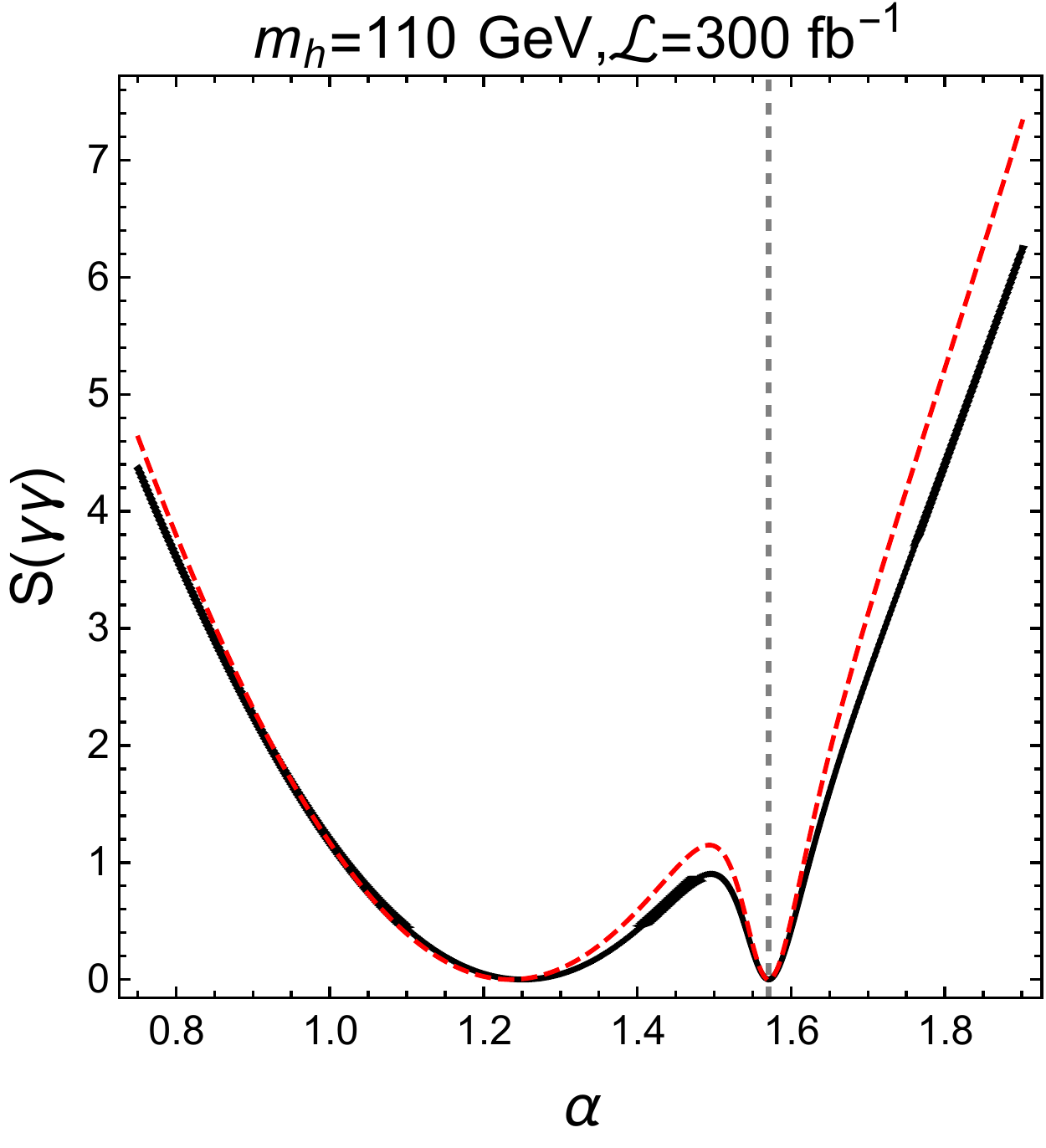}
\caption{\label{fig:chargedHiggs2}
In the left panel the allowed parameter space from the signal strength data is plotted with and without a charged Higgs. The blue contour shows the allowed 
parameter space without $H^\pm$ and the brown contour is with a $H^{\pm}$ with a mass of 
$200$ GeV. In the parameter region we are considering {\it i.e.}, $\tan\beta < 10$, the effect of adding a charged Higgs is minuscule. In the right panel we show the effect of the charged Higgs on the significance of observing $\gamma \gamma$
final state. The charged Higgs is found to enhance the significance for large values of $\sin(\beta-\alpha)$. The black (red dashed) line corresponds to the diphoton analysis without (with) the charged Higgs effects.
}

\end{center}
\end{figure} 

 \section{Fat-jet tagging techniques}
\label{sec:fatjet}
In this section we summarize the fat-jet tagging methods for Higgs and top-quark jets~\cite{Butterworth:2008iy,Plehn:2011tg}. 
We begin with the discussion on the reconstruction of a Higgs fat-jet. To start with, we combine all of the momentum four-vectors $(j_i)$
within $\Delta R = 0.8$ to form a fat jet ($J$) using the Cambridge-Aachen algorithm. The fat-jets
with $p_T > 200$ GeV are considered for further analysis.    
\begin{itemize}
\item The fat jet ($J$) is broken into two subjets ($j_{1}$ and $j_{2}$) and the heavier jet is labeled as $j_1$. 
\item The two subjets are considered if the mass of $j_1$ has a sufficient mass drop, i.e., $m_{j_1}< \mu m_{J}$ and the splitting between two jets defined as $y = \dfrac{{\rm min}(p_{T_1},p_{T_2})}{{\rm max}(p_{T_1},p_{T_2})}$ is greater than $y_{cut}$.\footnote{ This is to ensure not too asymmetric splitting between $j_1$ and $j_2$.}
This is a powerful cut to reduce the contaminations due to the QCD background.
We have considered $\mu =0.67$ and $y_{cut} = 0.09$ for our analysis~\cite{Butterworth:2008iy}. 

\item If the previous condition is not satisfied then $j_1$ is identified as $J$ and the procedure is repeated until both of the above conditions are satisfied.

\item The final jet is considered as the Higgs if both subjets are b-tagged and the mass of the filtered\footnote{To eliminate underlying events in the fatjet, it is filtered with $R_{\text{filter}}=0.3$ and three hard subjets are retained.} fat jet ($m_{J}$) is close to the Higgs mass.
\end{itemize}

Now we discuss the reconstruction of the top jet. We combine all of the momentum four-vectors $(j_i)$
within $\Delta R = 1.2$ to form a fat jet ($J$) using the Cambridge-Achen algorithm. The fat jets
with $p_T > 250$ GeV are considered for further analysis.    
\begin{itemize}
\item Inside a fat jet, a loose mass-drop criteria is employed such that $J \to j_{1}j_{2}$, $m_{j_2} < m_{j_1}$ and $m_{j_2} > 0.2 m_{J}$. The splitting takes place iteratively untill $m_{j_1} >$ 30 GeV. A fat jet is retained if it has at least three such subjets.
\item The three subjets are then filtered with $\Delta R = 0.3$ into five subjets. Only those fat jets with a total jet mass close to the top-quark mass are considered. The subjets which reconstruct the top mass are then reclustered into three subjets.
\item These subjets are then required to satisfy decay kinematics. Among the three pairs of invariant masses with these subjets, two of them are independent (as one of them satisfies $W$-mass criteria). In a two-dimensional space where the coordinates represent two independent invariant masses, top-like jets represent a thin triangular annulus, whereas the QCD jet is localized in the region of small pairwise invariant mass.
\end{itemize}

\section{Cross section}
\label{sec:csplot}

The dependences of the total cross section ($\sigma \times$ BR) on $\alpha$ and $\sin(\beta-\alpha)$ are listed in Table.~\ref{tab:crosssection-behaviour} and also displayed in Fig.~\ref{fig:cs2}. 

\begin{table}[htb]
\def\arraystretch{1.1}
\begin{center}
\begin{tabular}{|c|c|c|c|} \hline \hline
 & Total cross section  &  Parametric dependence   & Limit where the cross section vanishes \\ \hline
$\,\,$A$\,\,$ & $\sigma(p p \to h \to \gamma \gamma)$ & $\left(\dfrac{\cos\alpha}{\sin\beta}\right)^2 \times |\xi_h^\gamma|^2\times \dfrac{1}{\Gamma_h^{\text{tot}}}$ & $\alpha \to \pi/2$, $|\xi_h^\gamma|\to 0$  \\ \hline
%
$\,\,$B$\,\,$ & $\sigma(p p \to V h \to V \gamma \gamma)$ & $\sin(\beta-\alpha)^2 \times |\xi_h^\gamma|^2\times \dfrac{1}{\Gamma_h^{\text{tot}}}$ & $\alpha \to \beta $, $|\xi_h^\gamma|\to 0$ \\ \hline
$\,\,$C$\,\,$ & $\sigma(p p \to t \bar{t} h \to t \bar{t} b \bar{b} )$ & $\left(\dfrac{\cos\alpha}{\sin\beta}\right)^4 \times \dfrac{1}{\Gamma_h^{\text{tot}}}$ & $\alpha \to \pi/2$ \\ \hline
$\,\,$D$\,\,$ & $\sigma(p p \to V h \to V b \bar{b})$ & $\sin(\beta-\alpha)^2 \times \left(\dfrac{\cos\alpha}{\sin\beta}\right)^2  \times \dfrac{1}{\Gamma_h^{\text{tot}}}$ 
 & $\alpha \to \pi/2$, $\alpha \to \beta $  \\ \hline \hline
\end{tabular}
\caption{\label{tab:crosssection-behaviour} The dependences of the total cross section for various processes with respect to the coupling scale factors. The limits where the total cross section vanishes are also listed. The behavior of the total cross section for all four cases with respect to $\alpha$ for $\tan\beta=2$ and 6 is plotted in Fig.\ref{fig:cs2}.}

\end{center}
\end{table}
It can be easily seen from the expressions for the cases A and D listed in Table~\ref{tab:crosssection-behaviour} that the behavior of the total cross section for $pp \to h \to \gamma\gamma$ and $p p \to W h \to W b \bar{b}$ becomes identical with respect to $\alpha$ in the large-$\tan\beta$ regions. The same can also be verified from the $\tan\beta=6$ 
line in Fig.~\ref{fig:cs2}.

\begin{figure}[h] 
\begin{center}
\includegraphics[height=4.5cm,width=0.35\textwidth]{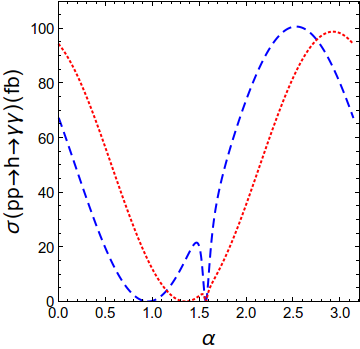}
\includegraphics[height=4.5cm,width=0.35\textwidth]{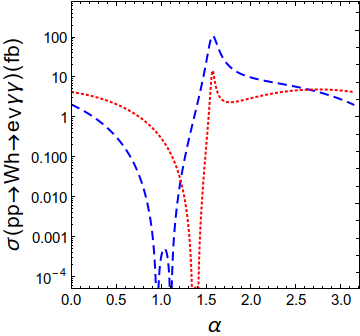}\\
\includegraphics[height=4.5cm,width=0.35\textwidth]{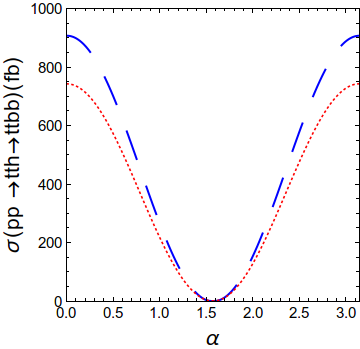}
\includegraphics[height=4.5cm,width=0.35\textwidth]{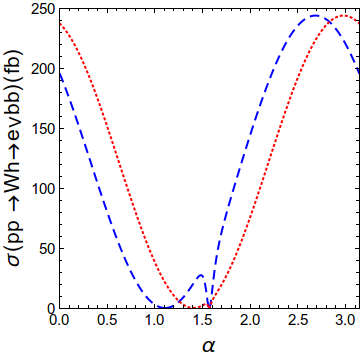}
\caption{A representative plot of ($\sigma \times$ BR) for light Higgs decaying to $\gamma \gamma$ and $b \bar{b}$ for $m_h=100$ GeV. The dashed line in blue corresponds to $\tan{\beta}=2$ while the dotted line in red corresponds to a $\tan\beta =6$.
}
\label{fig:cs2}
\end{center}
\end{figure}


\bibliography{references.bib}

\end{document}